\documentclass[prl,twocolumn,showpacs,preprintnumbers,amsmath,amssymb,superscriptaddress,aps,10pt]{revtex4-1}

\usepackage{graphicx}
\usepackage{dcolumn}
\usepackage{bm}
\usepackage{units}
\usepackage{subfigure}

\begin{document}

\title{Shear shuffling governs plastic flow in nanocrystalline metals:\\ An analysis of thermal activation parameters}

\author{M.~Grewer}
\email[]{m.grewer@mx.uni-saarland.de}
\affiliation{Experimentalphysik, Universit\"at des Saarlandes, Saarbr\"ucken, Germany}
\author{R.~Birringer}
\affiliation{Experimentalphysik, Universit\"at des Saarlandes, Saarbr\"ucken, Germany}

\date{\today}

\begin{abstract}
From strain rate- and temperature-dependent deformation studies on nanocrystalline PdAu alloys with grain sizes $\leq\unit[10]{nm}$, the shear activation volume ($\unit[6]{b^3}$), strain rate sensitivity ($0.03$) as well as the Helmholtz (\unit[0.9]{eV}) and Gibbs free energy of activation ($\Delta G = \unit[0.2]{eV}$) have been extracted. The close similarity to values found for metallic glasses indicates that grain boundary mediated shear shuffling dominates plasticity at the low end of the nanoscale. More fundamentally, we find that the energy barrier height exhibits universal scaling behavior $\Delta G \propto \Delta\tau^{3/2}$, where $\Delta\tau$ is a residual load, giving rise to a generalization of the Johnson-Samwer $T^{2/3}$ scaling law of yielding in metallic glasses.
\end{abstract}

\maketitle
\section{Introduction}
The good tensile ductility of conventional fcc polycrystalline metals relies on two essentials: presence and stress-induced multiplication of dislocations, which act as flow defect and propagate strain, in conjunction with the capability of strain or work hardening, which is basically due to intraplane dislocation interactions and self-organized dislocation-cell-structure formation \cite{Argon2008}. By contrast, in metallic glasses, dislocations are configurationally unstable and are consequently not available as carriers of strain. The missing long-range atomic order in metallic glasses favors the emergence of incipiently localized shear transformations (STs) upon loading \cite{Argon1979}. The concomitant shuffling or flipping of groups of atoms manifests as a flow defect, thus playing the same role as dislocations do in crystalline environments. With increasing applied load, STs typically self-organize in the form of shear bands, regions of high strain localization, which upon further increasing stress propagate through the material to eventually lead to catastrophic failure \cite{Schuh2007}. 

With regard to nanocrystalline (NC) metals \cite{Birringer1989}, it is obvious that none of these limiting cases applies. Clearly, the plasticity of NC metals involves a much higher degree of complexity for the following reasons: Since the volume fraction of grain boundaries (GBs) scales with the reciprocal grain size, the abundance of GBs at the nanometer scale supplies barriers for \textit{intergranular} slip transfer, and the nanometer-sized grains entail a reduced capacity of dislocation generation and intraplane dislocation interaction even at the upper limit of the nanometer scale of $\approx \unit[100]{nm}$. As a consequence, higher strength, lower activation volume, and higher strain-rate sensitivity have been observed \cite{Meyers2006,Dao2007}. Upon decreasing the grain size to the lower end of the nanoscale $\lesssim \unit[10]{nm}$, it is expected that \textit{intragranular} crystal plasticity is largely replaced by \textit{intergranular} plasticity, deformation processes that essentially emerge in the core regions of GBs \cite{Chokshi1989,Karch1987}. Computersimulations and experiments unraveled a variety of modes of plastic deformation related to GBs. So far the following processes have been identified: GB slip and sliding \cite{Vo2008,VanSwygenhoven2001,Weissmuller2011}, stress-driven GB migration coupled to shear deformation and grain rotation \cite{Cahn2006,Cahn2004,Legros2008}, as well as shear shuffling (ST) mediated plasticity \cite{Lund2005,Argon2006}, here operating in the confined space set up by the core region of GBs. Moreover, GB-ledges and triple junction lines, locations where typically three GBs meet, act as stress concentrators \cite{Gu2011,Asaro2005}, thereby effectively reducing the barrier for partial dislocation nucleation and emission. One of the intriguing aspects here is that in NC metals plastic deformation requires that those mechanisms operate together in the sense that deformation but also accommodation processes must at least partly coexist in order to make deformation happen in a compatible manner, thus avoiding brittle fracture and enabling substantial deformability. 

Therefore, one of the central remaining issues is to identify the relative importance of the variety of possible inter- and intragrain deformation modes and assign and quantify in which manner the relevant modes contribute with rising stress to overall strain. It is the aim of this work to deduce principal activation parameters for plasticity from experiment and to analyze these parameters to enable discriminating between the dominating mechanism(s) and the just possible mechanisms, here for NC $\mathrm{Pd}_{90}\mathrm{Au}_{10}$ in the limiting case of $D \lesssim \unit[10]{nm}$.
%
\begin{figure*}[t]
 \centering
     \subfigure[]{\includegraphics[width=0.40\textwidth]{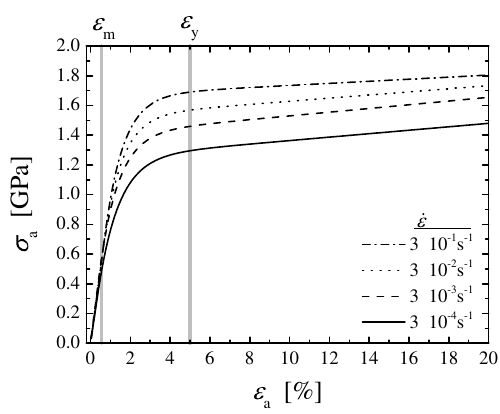}\label{fig:StressStrain}}
 \hspace*{1.3cm}
     \subfigure[]{\includegraphics[width=0.42\textwidth]{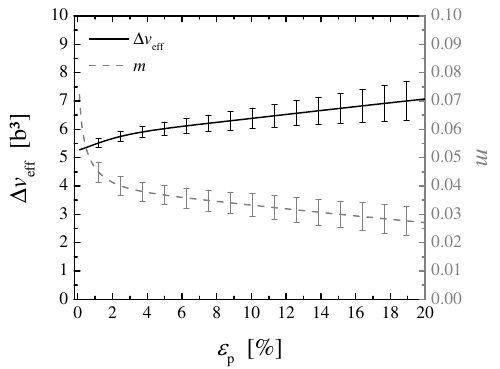}\label{fig:ActVol}}
 \caption[]{(a) Room temperature stress-strain curves of NC $\mathrm{Pd}_{90}\mathrm{Au}_{10}$ ($D\approx\unit[10]{nm}$) deformed under dominant shear at strain rates $\dot{\varepsilon}_{a}$ between $\unit[3\cdot 10^{-4}]{s^{-1}}$ and $\unit[3\cdot 10^{-1}]{s^{-1}}$. The gray markers indicate the strains corresponding to the deviation from linear elasticity ($\varepsilon_\mathrm{m}$) and the onset of yielding ($\varepsilon_\mathrm{y}$) respectively; the range between $\varepsilon_\mathrm{m}$ and $\varepsilon_\mathrm{y}$ is usually termed microplastic regime (for more details see \cite{Ames2012}). (b) Effective activation volume $\Delta v_\mathrm{eff}$ and strain rate sensitivity $m$ of NC Pd$_{90}$Au$_{10}$ as a function of plastic strain $\varepsilon_{p}$. Error bars are shown for every 20th data point.}
 \label{fig:ExperimentalData}
\end{figure*}
%

The activation parameters \cite{Argon2008} that are most informative for probing the mechanism(s) of thermally activated plasticity include: first of all, the Gibbs free energy of activation $\Delta G^*$, which at low temperature $T$, where entropic effects play a minor role, can be approximated by the enthalpy of activation $\Delta H^*$. In transition state theory \cite{Seeger1955}, the free-energy difference between the initial state ($i$) and the saddle point configuration ($s$) is usually denoted by $Q$ having the connotation of an activation free energy and thus $\Delta G^* \equiv Q$. Secondly, there is the shear activation volume $\Delta v^*_{\tau}$ as well as the phenomenological  strain rate sensitivity $m$, and thirdly, the activation dilatation $\Delta v^* _P$, which is particularly sensitive to the dependence of yield stress on hydrostatic pressure. The full set of activation parameters allows one to identify, discriminate, and/or obtain information about the relative importance of thermally activated deformation mechanisms. We concentrate in this study on the shear activation volume $\Delta v^*_{\tau}$, strain rate sensitivity $m$, and the Gibbs free energy of activation $\Delta G^*$. All of them are extracted for NC $\mathrm{Pd}_{90}\mathrm{Au}_{10}$ in the limiting case of $D \lesssim \unit[10]{nm}$; results on the activation dilatation $\Delta v^*_P $ will be communicated in a forthcoming report.

\section{Specimen preparation and mechanical testing}
NC $\mathrm{Pd}_{90}\mathrm{Au}_{10}$ specimens with average grain sizes $D \lesssim \unit[10]{nm}$ were prepared by inert gas condensation and subsequent consolidation \cite{Birringer1989}. The microstructure of these so-prepared samples is characterized by a lognormal grain size distribution \cite{Krill1998}, a random texture \cite{Markmann2003} and a GB-misorientation distribution that resembles the random MacKenzie distribution \cite{Schaefer2000}. It is dominated by high-angle GBs, which is also reflected by the area-weighted average GB energy of NC Pd $\gamma_A\approx \unit[0.8]{J/m^2}$ obtained from calorimetry \cite{Birringer2002}. For mechanical testing, miniaturized shear-compression specimens (SCSs) \cite{Rittel2002} were cut from as-prepared disk-shaped samples. Tests were carried out at room temperature and at $\unit[77]{K}$. Regarding sample preparation, mechanical testing, and data reduction, all relevant details and graphs can be found in \cite{Ames2010, Ames2012}. To get rid of corrections for machine stiffness etc., we meanwhile improved the accuracy of the displacement measurement by utilizing a high resolution optical set-up consisting of a Zeiss SteReo Discovery V12 optical microscope (ca. $100\times$ magnification) and a HighSpeedStar 3G CMOS camera capable of recording up to 1000 images per second at \unit[1]{MP} resolution. The recorded images were processed by digital image correlation using the software package Lavision DaVis 7.2 to deduce sample displacement. Von Mises equivalent stress and strain values are computed by the finite element method (FEM) using Abaqus \cite{Ames2010}. In Fig. \ref{fig:StressStrain}, we display the so-obtained stress-strain curves reflecting a slope of $\unit[105]{GPa}$ in the elastic regime that is in good agreement with the high frequency Youngs modulus of $\unit[110]{GPa}$ deduced from ultrasound measurements \cite{Grewer2011}.

\section{Shear activation volume: concept and determination}
We first analyze how the measured or effective shear activation volume $\Delta v_{\mathrm{eff}}$ is related to experimentally accessible quantities (applied stress and strain rate). We start with the definition of the shear activation volume $\Delta v^*_{\tau _a}$ that is assumed to be associated with a unique deformation mechanism \cite{Argon2008} and given as
\begin{equation}
\Delta v^*_{\tau _a} = \left. -\frac{\partial\Delta G^*(\tau_{a},\hat{\tau})}{\partial \tau_{a}}\right\vert_{T,P}, \label{eq:energyact}
\end{equation}
where $\Delta G^*$ must be supplied by thermal fluctuations at constant applied stress $\tau _a$, pressure $P$, and temperature $T$ to reach a saddle point configuration of the shear barrier and thus cause an activated process to take place. The shear resistance $\tau$ is a material property and is defined as $\tau = V^{-1}\, \partial F/\partial\gamma$, where $\gamma$ denotes the shear strain, $V$ is the volume of the system, and $F$ is the Helmholtz potential. In particular, $\hat{\tau}$ is the athermal (rate-independent) threshold stress characterizing the maximum level of shear resistance as $T \rightarrow 0$. The following identity holds for the Helmholtz free energy of activation: $\Delta F^* = \Delta G^* + \Delta W^*$, where $\Delta W^* = V \tau _a \, \Delta\gamma_c$ denotes the external mechanical work supplied during activation, and $\Delta\gamma_c$ is the activation strain related to localized inelastic shear events with $\Delta\gamma_c = \gamma_s -\gamma_i$. We note that $\Delta v^*_{\tau _a}$ is an apparent activation volume that is given to first order by $\Delta v^*_{\tau _a} = \Omega \Delta\gamma_c$, where $\Omega$ is the true volume of a cluster of atoms that has been involved in a local shear event. For applied stresses close to $\hat{\tau}$, where the reverse rate of deformation can be neglected, the inelastic net strain rate 
$\dot{\gamma_p}$ \cite{Argon2008} is given by 
\begin{equation}
\label{eq:strainrate} 
\dot{\gamma_p} = \dot{\gamma}_0 \exp[-\Delta G^*(\tau _a, \hat{\tau})/kT], 
\end{equation}
and the preexponential factor $\dot{\gamma}_0$ represents a reference strain rate. It essentially depends on the volume fraction of fertile sites that trigger configurational transformations, the unconstrained transformation shear strain, and a normal mode frequency of atom clusters of size $\Omega$ taking part in configurational changes along the activation path. Usually, $\dot{\gamma}_0$ is considered constant, an assumption we also made for the sake of feasibility when deriving Eq. \ref{eq:actvol} in the next paragraph. This assumption may fail, however, when the preexponential term becomes stress-dependent. In such a scenario, we expect (a) stress-dependent correction term(s) to $\dot{\gamma}_0$ that sensitively depend(s) on the deformation mechanism(s) and  atomistic and microstructural details of the material. As a consequence, $\Delta v^*_{\tau _a}$ will also change its behavior.   
    
With the continuum theory of plasticity, we find the kinetic shear rate $\dot{\gamma_p}$ related to the tensile strain rate $\dot{\varepsilon}_p$ by $\dot{\gamma_p} = \sqrt{3}\, \dot{\varepsilon}_p$, and the analog regarding shear and tensile stress reads $\tau_a = \sigma_a / \sqrt{3} $ \cite{Argon2008}. Solving Eq. \ref{eq:strainrate} for $\Delta G^*$ and substituting $\dot{\gamma_p}$ by $\dot{\varepsilon}_p$, we derive from Eq. \ref{eq:energyact}
\begin{equation}
\Delta v^*_{\tau_{a}} = \sqrt{3} \, kT\left.\left(\frac{\partial\ln\dot{\varepsilon}_p}{\partial\sigma_{a}}\right)\right\vert_{T,P,\varepsilon_{p}} , \label{eq:actvol}
\end{equation}
where $\dot{\varepsilon}_p $ is related to the prescribed strain rate $\dot{\varepsilon}_a$ by $\dot{\varepsilon}_p /\dot{\varepsilon}_a = (1 - \Theta/C)$ where $\Theta$ is the tangent modulus, and $C$ denotes the effective Young´s modulus given by the slope of the linear part of the applied stress-strain curve \cite{Ames2012,Saada2011}. Clearly, the applied stress $\sigma_a$ is a function of the applied strain and strain rate $\sigma_a = \sigma_a (\varepsilon_a,\dot{\varepsilon}_a)$, and since the concept of shear activation volume is related to inelastic deformation, it follows that $\Delta v^*_{\tau _a}$ has to be determined at a given plastic strain $\varepsilon_p$. We would like to point out here that values of $\Delta v^*_{\tau _a}$ derived according to Eq. \ref{eq:actvol} from experimental data have to be interpreted as effective shear activation volume $\Delta v^{*,\mathrm{eff}}_{\tau _a}$ since it is \textit{a priori} not known which mechanisms contribute in which manner to overall deformation. In the following we use the short notation $\Delta v^{*,\mathrm{eff}}_{\tau _a} \equiv \Delta v_{\mathrm{eff}}$. The phenomenological strain rate sensitivity $m$ is defined as $m = ( \partial \ln \sigma_a / \partial \ln \dot{\varepsilon}_a) \arrowvert_{T,P,\varepsilon_{a}}$ and also has the character of an effective quantity.
%
\begin{figure*}[t]
 \centering
     \subfigure[]{\includegraphics[width=0.35\textwidth]{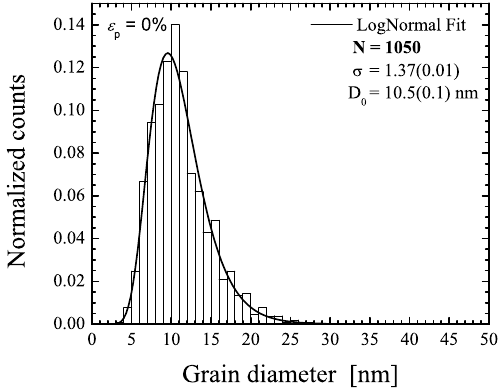}\label{fig:TEM0}}
 \hspace*{0.5cm}
     \subfigure[]{\includegraphics[width=0.35\textwidth]{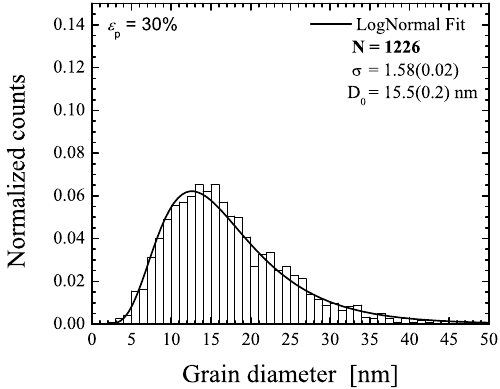}\label{fig:TEM30}}
 \hspace*{0.5cm}
     \subfigure[]{\includegraphics[width=0.2\textwidth]{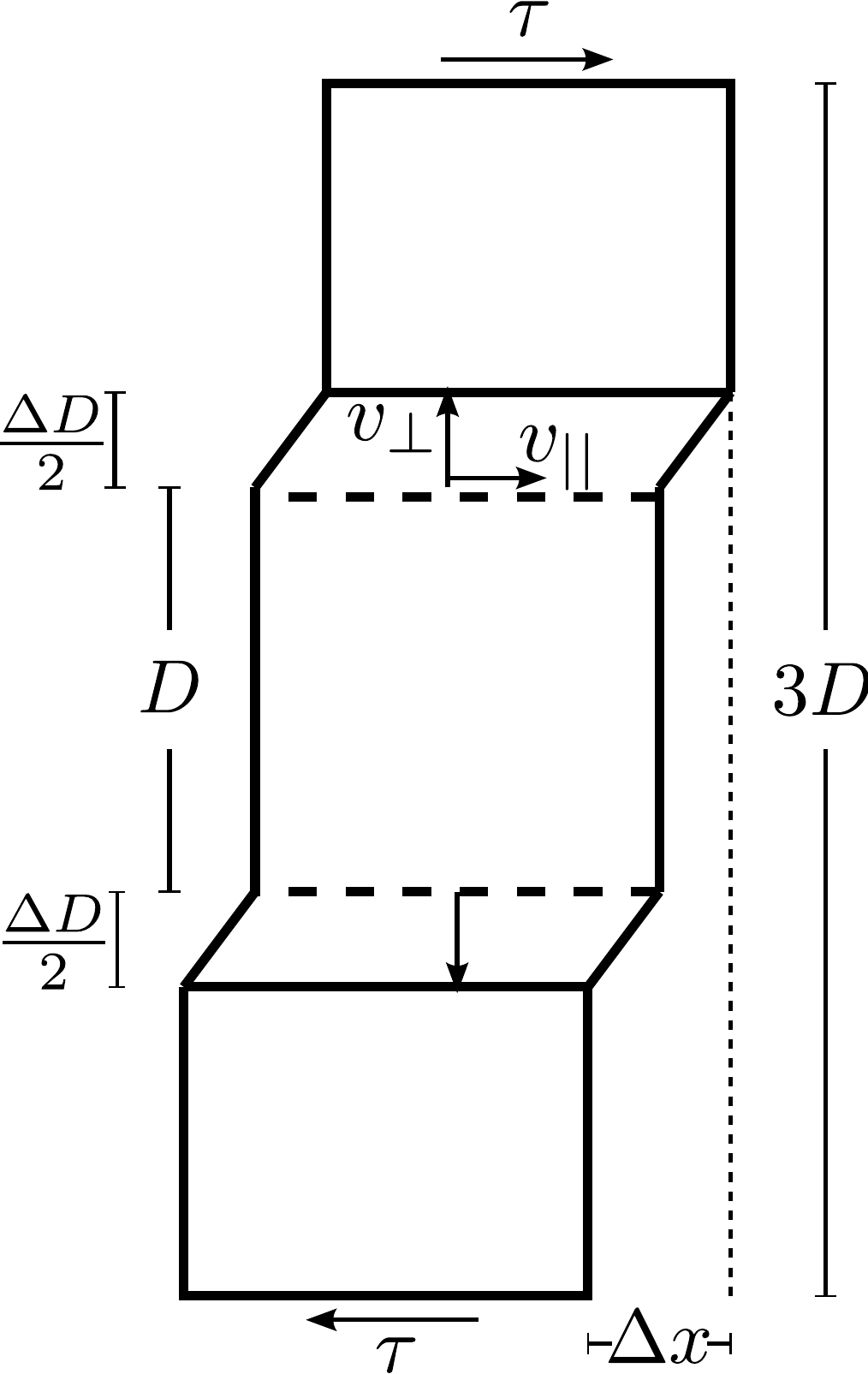}\label{fig:Coupling}}
 \caption[]{(a) Grain size distribution of undeformed NC Pd$_{90}$Au$_{10}$ derived from TEM dark field images. The solid line is a lognormal fit to the histogram with width $\sigma$, median $D_0$ and $N$ denotes the number of counted grains. (b) Grain size distribution after deformation to \unit[30]{\%} plastic strain. (c) Sketch of a three grain column where the middle grain exhibits stress driven GB migration in response to $\tau$. For more details, see the text.}
 \label{fig:SDGBM}
\end{figure*}
%

\section{Shear activation volume: results and discussion}
The evolution of $\Delta v_{\mathrm{eff}}$ as a function of plastic strain is displayed in Fig. \ref{fig:ActVol}; we note that values for $\Delta v_{\mathrm{eff}}$ are by a factor of $\sqrt 3$ larger than the ones given in \cite{Ames2012}, where tensile stress has been used in the definition rather than shear stress. The evolution of the strain-rate sensitivity $m$ is also shown in Fig. \ref{fig:ActVol}; we take $m$ at fixed values $\varepsilon_{p}$ to allow direct comparison with $\Delta v_{\mathrm{eff}}$. 

Fundamentally, the activation volume for crystalline metals is bounded by $\approx \unit[10^3]{b^3}$ ($\approx 2 \cdot 10^1 \mathrm{{nm}^3}$ ) when forest dislocation cutting dominates plasticity \footnote{We assumed a value of $\unit[0.275]{nm}$ for the magnitude of the burgers vector $\mathrm{b}$} and on the lower end by $\approx \unit[0.02-0.1]{b^3}$ ($\approx \unit[2 \cdot 10^{-3}- 4 \cdot 10^{-4}]{{nm}^3}$) which is indicative of creep processes \cite{Frost1982}. They are based on point defect migration and essentially constitute deformation modes such as Nabarro-Herring \cite{Herring1950} and Coble \cite{Coble1963} diffusion creep, usually manifesting the concomitant phenomenon of GB sliding \cite{Langdon2006}. Based on the magnitude of $\Delta v_{\mathrm{eff}} \approx 6 \mathrm{b^3}$ extracted for NC $\mathrm{Pd}_{90}\mathrm{Au}_{10}$ it can be concluded that intragranular lattice dislocation activity and creep processes can be ruled out as contributing in an appreciable manner to strain propagation by virtue of a discrepancy in magnitude of their respective $\Delta v^*_{\tau _a}$ values. This view is also supported by the values found for $m$ which are more than a magnitude smaller than typical values for Coble creep ($m \approx 1.0 $) \cite{Coble1963} and GB sliding ($m \gtrsim 0.3$) \cite{Luthy1979,Langdon2006}. 

There is a misuse of nomenclature in the pertinent literature dealing with nanoplasticity through  associating GB-mediated deformation with GB sliding. Originally, the term GB sliding was introduced to denote the rigid body translation of abutting crystallites along a shared interface that produces offsets in marker lines at the GBs. There are two different modes of GB sliding: Rachinger sliding \cite{Rachinger1952}, which must be accommodated by intragranular dislocation glide and climb, and Lifshitz sliding \cite{Lifshitz1963a}, which is based on stress-directed diffusion of vacancies and is self-accommodating. Since both types of GB sliding occur under creep conditions, they are observed at elevated temperatures, and not until a crossover temperature of $\approx \unit[0.5]{T_m}$ has been reached \cite{Cahn2006}. Moreover, as it is nearly impossible to experimentally identify and quantify GB sliding in nanoscale microstructures or discriminate the so-called GB sliding from the evolution of shear transformations (STs) and successive avalanches of STs \cite{Argon2006a,Demkowicz2005,Dahmen2009,Maloney2006}, we decided to refer in the following to the more general concept of ST-mediated plasticity \cite{Argon2013}, which bears a resemblance to events of self-organized criticality \cite{Jensen2000}, but may also lead to marker line shifts at GBs without requiring creep conditions. 

Quite generally, we may argue that the high stress levels present in our experiment at room temperature suggest that shear mechanisms should almost instantly overtake processes of diffusional matter transport. As a result, likely mechanisms that may be incorporated in $\Delta v_{\mathrm{eff}}$ are related to partial dislocation activity (PDA) involving nucleation and glide, shear-stress driven GB migration (SDGBM) including grain rotation and generation of shear strain, as well as ST-mediated plasticity. In the following paragraph, we discuss how $\Delta v_{\mathrm{eff}}$ can be decomposed into contributions related to the above-mentioned mechanisms.

Assuming additivity of strains resulting from the PDA, SDGBM, and ST mechanisms, we can write for the effective plastic strain rate, $\dot{\gamma}_{\mathrm{eff}}$, 
\begin{equation}
\label{eq:ratebalance}
\dot{\gamma}_{\mathrm{eff}} = \dot{\gamma}_{\mathrm{PDA}} + \dot{\gamma}_{\mathrm{SDGBM}} + \dot{\gamma}_{\mathrm{ST}}.  
\end{equation}
Rewriting Eq. \ref{eq:actvol} as $\Delta v_{\mathrm{eff}} = kT (\partial \ln\dot{\gamma}_{\mathrm{eff}}/ {\partial\tau_{a}})$, substituting $\dot{\gamma}_{\mathrm{eff}}$ by Eq. \ref{eq:ratebalance} and using Eq. \ref{eq:strainrate}, it is straightforward to derive
\begin{equation}
\label{eq:veffective}
\Delta v_{\mathrm{eff}} = \sum_{i} \left(\frac{\dot{\gamma}_i}{\dot{\gamma}_{\mathrm{eff}}}\right) \Delta v^*_i , 
\end{equation} 
where the summation runs over $i =$ PDA, SDGBM, and ST. The expression in the round bracket obeying $\sum_{i} (\dot{\gamma}_i / \dot{\gamma}_{\mathrm{eff}}) = 1 $ emerges as the shares of the different mechanisms to overall deformation. In what follows, we argue that the share of SDGBM to overall strain plays a minor role, and as a result Eq. \ref{eq:veffective} can be simplified.

To identify and characterize SDGBM that may occur during deformation, we applied focused ion beam (FIB) to prepare thin lamellae from the undeformed part of the specimen and the gauge section having experienced $30 \%$ plastic strain. We then took transmission electron microscopy (TEM) dark field images that have been analyzed in terms of size histograms, which are shown in Fig. \ref{fig:SDGBM}. The shift of the median $D_0$ from $\unit[10.5]{nm}$ to $\unit [15.5]{nm}$ clearly indicates that SDGBM took place during deformation. Referring to a simple model shown in Fig. \ref{fig:Coupling}, we estimate the share of SDGBM to overall strain. 

The shear strain $\gamma_{\mathrm{SDGBM}}$ induced through the migration of GBs in response to the applied shear stress $\tau$ is given by $\gamma_{\mathrm{SDGBM}} = \Delta x / 3 D$. Using an average coupling factor of $ \langle \beta \rangle \approx 0.3$ \cite{Molodov2011} to describe the ratio of migration of both boundaries parallel ($v_{||}$) and perpendicular ($v_\bot$) to the applied stress, we find $\gamma_{\mathrm{SDGBM}} = \beta \Delta D / 3 D \approx0.05$, where $\Delta D$ is set equivalent to the increase of the median $\Delta D_0 = \unit[5]{nm}$. With $\varepsilon = \gamma / \sqrt{3}$, SDGBM contributes \unit[2.8]{\%} plastic strain or roughly $1/10$th to the overall deformation at \unit[30]{\%} plastic strain. 

For the sake of simplicity, we neglect the SDGBM-contribution in Eq. \ref{eq:veffective}. With the value $\Delta v_{\mathrm{eff}} \approx 6 \mathrm{b^3}$ determined for NC $\mathrm{Pd}_{90}\mathrm{Au}_{10}$, the theoretical value for $\Delta v^*_{\mathrm{PDA}} \arrowvert_{\unit[10]{nm}} \approx 10 \mathrm{b}^3$ \cite{Gu2011}, and $\Delta v^*_{\mathrm{ST}} \approx 5 \mathrm{b}^3$ derived from experiment \cite{Pan2008,Ju2011}, we find $(\frac{\dot{\gamma}_{\mathrm{ST}}}{\dot{\gamma}_{\mathrm{eff}}}) \approx 4\, (\frac{\dot{\gamma}_{\mathrm{PDA}}}{\dot{\gamma}_{\mathrm{eff}}})$ so suggesting that ST-mediated plasticity dominantly contributes to overall strain. At first sight, this result seems to be in contradiction to the usual assumption that PDA-based deformation controls the overall deformation behavior of NC metals even below $\unit[20]{nm}$ grain size. This conflict can be reconciled by noting a recent work on texture formation that is capable of accummulating and storing information of subtle changes in microstructure due to dislocation-based plasticity. Studying texture formation in NC $\mathrm{Pd}_{90}\mathrm{Au}_{10}$ induced by high-pressure torsion, Skrotzki et. al. \cite{Skrotzki2013} find that for grain sizes below $\unit[20]{nm}$ and applied strains up to $\gamma \approx 1$ any texture formation is missing.  Moreover, they observed that twinning and stacking fault formation is basically absent, and it is argued that due to the extremely low remnant dislocation density in individual nanograins cross-slip and recovery by edge dislocation climb is unlikely. At strains $\gamma > 1$ texture formation starts evolving and originates mainly from slip of $1/6 \langle 112 \rangle$ partial dislocations that are nucleated from GBs and glide on \{111\} planes. Since our experiments entail strains $\gamma \leq 0.4$, it appears safe to follow our assertion of negligible strain contribution of dislocation activity to overall strain. In the following we are going to scrutinize our conjecture of dominating ST-mediated plasticity by determining activation energies, $\Delta G _{\mathrm{eff}}$ and $\Delta F _{\mathrm{eff}}$, and comparing them with available data for metallic glasses.

\section{Activation energy: Concept and determination}
Before discussing experimental details, we first look into how $\Delta G _{\mathrm{eff}}$ relates to the activation energies of the involved mechanisms. Since the form of Eq. \ref{eq:strainrate} applies to the effective activation energy $\Delta G _{\mathrm{eff}}$ as well as to the specific activation energies $\Delta G^*_i$ of possible mechanisms, it follows from combining Eqs. \ref{eq:strainrate} and \ref{eq:ratebalance}
\begin{equation}
\label{eq:deltaG}
\Delta G_{\mathrm{eff}} = -kT\, \ln \left[\sum_{i} \left(\frac{\dot{\gamma}_{0i}}{\dot{\gamma}_{0, \mathrm{eff}}}\right) \exp {\left(\frac{- \Delta G^*_i}{k
T}\right)}\right]. 
\end{equation}
In contrast to $\Delta v_{\mathrm{eff}} $, which entails a weighted linear superposition of the possible mechanisms (Eq. \ref{eq:veffective}), the effective activation energy $\Delta G _{\mathrm{eff}}$ is given as the logarithm of the sum of exponential terms. Even for small differences in the $\Delta G^*_i$-terms of conceivable mechanisms, it is evident that at room temperature the smallest value of $\Delta G^*_i \approx \Delta G _{\mathrm{eff}}$. We anticipate that the value $\Delta G _{\mathrm{eff}}$ derived from experiment should agree with some characteristic value $\Delta G^*_i$ obtained for ST-based plasticity in metallic glasses.

A suppostion made here is that ST-mediated plasticity can take place in an unconstrainded manner so that coupling to e.g. accomodation modes having significantly higher $\Delta G^*_i$ values is missing. In fact, the argument that the minimal $\Delta G^*_i$ approximates the measured $\Delta G _{\mathrm{eff}}$ presumes undisturbed superposition of deformation modes making up for the overall strain. From a mechanistic point of view, it is implied that the contiguous network of GBs, the core regions of which are characterized by atomic mismatch associated with excess volume \cite{Birringer1995,Krill2001}, offers a plethora of fertile sites, particularly at the low end of the nanoscale. They are predestined to trigger local shear events, most notably since the shear stiffness in those core regions is reduced by about $30\%$ compared to the abutting crystal lattices \cite{Grewer2011}, and so the overall deformation may evolve without requiring any significant deformation share originating from generation of intracrystalline strain increments.       

\begin{figure}[h]
	\centering
		\includegraphics[width=0.36\textwidth]{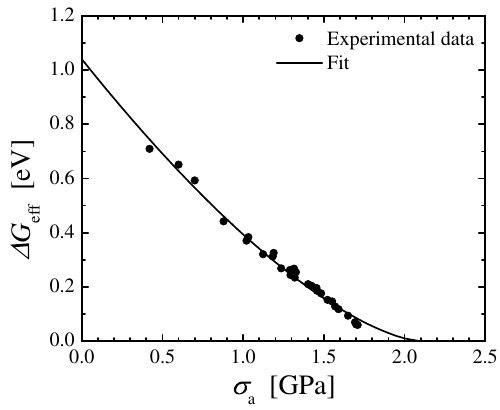}
	\caption{Gibbs activation enthalpies of NC Pd$_{90}$Au$_{10}$ plotted as a function of applied stress $\sigma_a$. The shown data represent the dependency of $\Delta G$ on applied strain rate at room temperature [Fig. \ref{fig:StressStrain}] and \unit[77]{K} \cite{Ames2012}. The full line is a fit of Eq. \ref{eq:powerlaw} to the data (continuum theory of plasticity \cite{Argon2008}: $\tau = \sigma/\sqrt{3}$).} 
	\label{fig:ActEnthalpy}
\end{figure}

For determination of $\Delta G_{\mathrm{eff}}$, we refer to Eq. \ref{eq:strainrate} to find $\Delta G_{\mathrm{eff}}(\hat{\tau}, \tau_a) = kT\, [\ln({\dot{\gamma}_{0, \mathrm{eff}}}/{\dot{\gamma}_{\mathrm{eff}}})]$. Experimentally, we have direct access to $\sigma_a = \sigma_a(T, \dot{\varepsilon}_{p, {\mathrm{eff}}} \equiv \dot{\varepsilon}_{{\mathrm{eff}}} )$ and using continuum plasticity also to $\tau_a = \tau_a (T, \dot{\gamma}_{\mathrm{eff}}) $. Extracting values for $\Delta G_{\mathrm{eff}}$ therefore requires to express how $\Delta G_{\mathrm{eff}}$ depends explicitly on $\hat{\tau}, \tau_a$. At stresses in the vicinity of $\hat{\tau}$ (this case), theory predicts \cite{Cahn2001,Cottrell2002} that the thermal fluctuation energy for activation of a local shear event has the form of a power-law $\Delta G_{\mathrm{eff}} = C (\hat{\tau} - \tau_a)^n\,$, where $C$ is a constant and the exponent may assume values of $n = 3/2$ or $2$. When $\tau_a \rightarrow \hat{\tau}$, it follows that $\Delta G_{\mathrm{eff}}(\hat{\tau}, \tau_a) \rightarrow 0$ which describes the transition from thermally activated to athermal deformation. In this case the initial state and the saddle point configuration of the shear barrier merge thus implying that $\Delta v_{\mathrm{eff}} \rightarrow 0$ with $\tau_a \rightarrow \hat{\tau}$. In other words, $\Delta G_{\mathrm{eff}}(\hat{\tau}, \tau_a)$ must approach zero value at $\tau_a = \hat{\tau}$ with zero slope  and necessarily $n>1$. It is straightforward to express the power-law in the form
\begin{equation}
\label{eq:powerlaw}
\Delta G _{\mathrm{eff}} = \Delta F_0\, [1- (\tau_a / \hat{\tau})]^n 
\end{equation}
where $\Delta F_0$ is the Helmholtz potential energy barrier at zero stress, which is a material property \cite{Bulatov1994b,Eshelby1957} for a given deformation mechanism. Using continuum plasticity, we derive $\sigma_a = \hat{\sigma}\, [1 - (\Delta G_{\mathrm{eff}} / \Delta F_0 )^{1/n}]$ with $\Delta G_{\mathrm{eff}} = kT \ln(\dot{\varepsilon}_{0,_{\mathrm{eff}} }/ \dot{\varepsilon}_{\mathrm{eff}})$. The value of $\dot{\varepsilon}_{0,_{\mathrm{eff}} }$ is\textit{ a priori} unknown, however, the latter equation stipulates that a continuous and unique curve must exist for data points when $\sigma_a$ is plotted versus $\Delta G_{\mathrm{eff}}$. The central task is to find an optimum value(s) for $\dot{\varepsilon}_{0,\mathrm{eff}}$ that fulfills this requirement within the lowest scatter of data points \cite{Kocks2003}. In Fig. \ref{fig:ActEnthalpy}, we display the so-obtained data points for $\Delta G_{\mathrm{eff}}(\sigma_a)$ over a broad range of experimentally accessible stress values.
%
%

\section{Activation energy: results and discussion}
In order to extract values for the yet unknown parameters $\hat{\sigma}$, $\Delta F_0$ and $n$, we performed a least-squares fit to the full set of data points using Eq. \ref{eq:powerlaw}. Assuming a fixed exponent $n = 3/2$ we find for the free parameters $\Delta F_0 = \unit[1.04]{eV}$ and $\hat{\sigma} = \unit[2.1]{GPa}$. The fit based on these values is shown as a full line in Fig. \ref{fig:ActEnthalpy}. Treating $\Delta F_0$, $\hat{\sigma}$ as well as $n$ as free parameters, we find that the exponent $n$ depends sensitively on small variations $\delta \hat{\sigma}$ of the order of $\unit[0.05]{GPa}$. However, we can independently estimate $\hat{\sigma}$ by arguing that the strain-rate dependent onset stress of inelastic deformation should approximate $\hat{\sigma}$ when $1 / \dot{\varepsilon}_a \rightarrow 0$. Thus using the experimentally extracted value $\dot{\varepsilon}_{0,_{\mathrm{eff}} } = \unit[10^8]{s^{-1}}$  at the onset of inelastic deformation as an upper bound for $\dot{\varepsilon}_a$, we find $\hat{\sigma} = \unit[2.05]{GPa}$ by linear extrapolation. A least-squares fit to the data points in Fig. \ref{fig:ActEnthalpy} based on Eq. \ref{eq:powerlaw} now with $\Delta F_0$ and $n$ as free parameters and fixed $\hat{\sigma} = \unit[2.05]{GPa}$ yields $\Delta F_0 = \unit[1.0]{eV}$ and $n = 1.4$. The good agreement between both approaches makes us feel confindent that the extracted parameter values are meaningful. We come back to a physical interpreatation of the $n = 3/2$ power-law behavior (Eq. \ref{eq:powerlaw}) in a later paragraph but now concentrate on activation energies.              

\begin{figure}[h]
	\centering
		\includegraphics[width=0.36\textwidth]{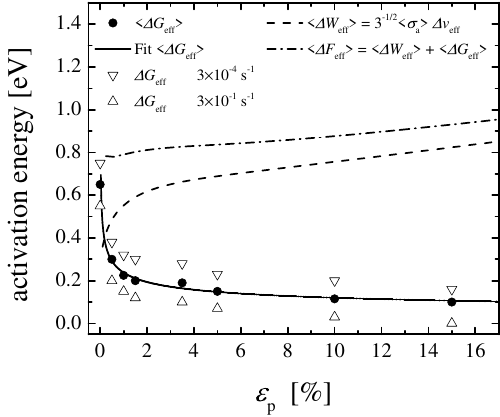}
	\caption{Average effective activation energies $\langle\Delta G_\mathrm{eff}\rangle$, $\langle\Delta W_\mathrm{eff}\rangle$ and $\langle\Delta F_\mathrm{eff}\rangle$ as a function of plastic strain. $\langle\Delta G_\mathrm{eff}\rangle$ (black dots) is taken to be the arithmetic mean of the energies $\Delta G_\mathrm{eff}(\dot{\varepsilon}_a)$, where the open triangles are associated with the applied minimal and maximal strain rates.}
	\label{fig:Activation}
\end{figure}

Since $\Delta G_{\mathrm{eff}}$ depends on $\dot{\varepsilon}_{\mathrm{eff}}$ and $\sigma_a$ is related to $\varepsilon_{p, \mathrm{eff}} \equiv \varepsilon_p $ via the stress-strain curve, we display how $\Delta G_{\mathrm{eff}}$ varies with $\varepsilon_p$ and $\dot{\varepsilon}_{\mathrm{eff}}$ in Fig. \ref{fig:Activation}. To compare with literature data, we also compute $\Delta F_{\mathrm{eff}} = \Delta G _{\mathrm{eff}} + \Delta W_{\mathrm{eff}}$, where the mechanical work done by the external agency is defined as $\Delta W = V \tau_a \Delta\gamma $. It is plausible to substitute $ V \Delta\gamma $ by the local quantities of the flow defect $\Omega \Delta \gamma_c = \Delta v_{\mathrm{eff}}$, thus obtaining $\Delta F_{\mathrm{eff}} = \Delta G_{\mathrm{eff}} + (\sigma_a/\sqrt{3})\, \Delta v_{\mathrm{eff}}(\varepsilon_p) $, which is, together with $\Delta W_{\mathrm{eff}}$, also shown in Fig. \ref{fig:Activation}. In the regime of microplasticity, it is clearly reflected how increasing mechanical work reduces $\Delta G_{\mathrm{eff}}$ to then reach a nearly plateau behavior for macroscopic plastic flow, which is characterized by $\Delta G_{\mathrm{eff}} \approx \unit[0.15]{eV}$ and $\Delta F_{\mathrm{eff}} \approx \unit[0.9]{eV}$. It emerges that $\Delta F_{\mathrm{eff}}\arrowvert_{\varepsilon_p = 0} = \unit[0.78]{eV}$ is smaller than $\Delta F_0$, extracted from the fit in Fig. \ref{fig:ActEnthalpy}, indicating the expected Helmholtz potential difference between the initial stress-free state and the still stable but stressed state at $\varepsilon_p = 0$. We now compare this data set with data obtained for inelastic deformation of metallic glasses. 
%
%

Studying yielding and plastic flow of a Zr-based metallic glass \cite{Klaumunzer2010}, it has been argued that the activation of STs is the rate limiting step of thermally activated shear-band propagation at a macroscopic scale. Hence temperature-dependent shear displacement jump velocity measurements have been performed to deduce an activation energy of $\Delta G^* = \unit[0.32]{eV}$. 

The investigation of room-temperature anelastic relaxation behavior has been exploited to characterize the properties of STs in Al-rich metallic glasses \cite{Ju2011}. An activation energy distribution of $0.85 < \Delta F^*_m < \unit[1.26]{eV}$ has been extracted, where the index $m$ relates to the number of atoms involved in a cluster undergoing a ST. It has been estimated that the size of these ST-events, $\Omega$, entails between 15-20 atoms. This result is based on the reasonable assumption $\Delta\gamma^*_c \approx 0.2$ \cite{Argon2008,Ju2011}. Dividing our experimentally determined $\Delta v_{ \mathrm{eff}}$ by $0.2$, we find a remarkably similar $\Omega \approx 32$ atoms in NC $\mathrm{Pd}_{90}\mathrm{Au}_{10}$. 

Atomic scale simulations have been used to study the plastic event distributions in the plastic flow state of a Lennard-Jones CuZr metallic glass \cite{Rodney2009}. By exploring the potential energy landscape the activation energy spectrum has been derived from unloading at different shear stress levels. In the limit of high stresses the most probable activated states are characterized by an activation energy (not further specified) of $\unit[0.35]{eV}$, and the spectrum of energies between $\unit[0.05]{eV}$ and $\unit[0.35]{eV}$ is populated with a frequency which is reduced by a third compared to the most probable state.

Molecular-dynamics (MD) simulations based on a highly realistic embedded atom method (EAM) potential for a CuTi model glass were carried out to study the highly localized as well as spatial and temporal heterogeneous flow occurring in STs to find a minimal activation energy of $\Delta G^* \approx \unit[0.35]{eV}$ for viscous flow of successive shear events involving $\approx 140$ atoms \cite{Mayr2006}. 

Overall, there is a reasonable agreement between activation energies reported in the pertinent literature and the values found in this study, particularly for the onset of macroyielding that occurs at $\varepsilon_p \approx 3 \%$. For the Al-rich glass, there is even a surprisingly good accordance between both the activated cluster size and the Helmholtz activation energy. The overall slightly smaller values of activation energies of  NC $\mathrm{Pd}_{90}\mathrm{Au}_{10}$ may reflect the fact that transient dilatancy going along with STs is easier to accomplish in GBs due to preexisting enhanced free volume stored in the GB core regions (GB excess volume) \cite{Krill2001,Birringer1995}.   

On the basis of this evidence, it is suggested that ST-mediated deformation dominates the plastic flow of NC $\mathrm{Pd}_{90}\mathrm{Au}_{10}$ with an average grain size of $\approx \unit[10]{nm}$. We note that the shear modulus of GBs in NC metals is reduced by about 30\% compared to the respective bulk value \cite{Grewer2011} and therefore local shear shuffling is predestined to take place in the contiguous network of GBs. Since we have estimated the size of an ST to $\Omega\approx \unit[32]{atoms}$ in Pd$_{90}$Au$_{10}$, it is implied, based on the structural (or polyhedral) unit model of GBs \cite{Sutton1983}, that two or three structural units are involved in the rearrangements of a ST. Those structural units carry different amounts of excess volume depending on the degree of local misfit and therefore may act as fertile sites that trigger the ST. Moreover, the various structural units formed in relaxed GBs resemble the polyhedral building blocks of the liquid structure \cite{Bernal1964,Spaepen1975,Howe1997}.

In contrast to metallic glasses, the topology and connectivity of the areal defects (GBs) in NC metals seems to avoid macroscopic shear band formation, which would give rise to stick-slip dynamics, which is clearly not observed. Mechanistically, the self-organized arrangement of STs to form a shear band seems to be effectively impeded in NC metals through possible local bifurcation instabilities at triple junctions, which in turn inhibit strain localization and catastrophic failure.

\section{Universal scaling}
%
\begin{figure}[b]
	\centering
		\includegraphics[width=0.42\textwidth]{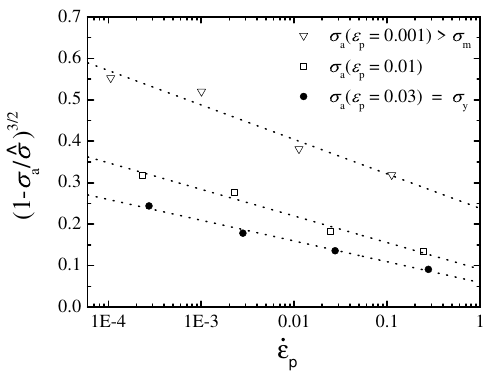}
	\caption{$(1-\sigma_\mathrm{a}/\hat\sigma)^{3/2}$ as a function of $\dot\varepsilon_\mathrm{p}$. To allow for easy reconstruction of the actual $\dot\varepsilon_\mathrm{p}$ values, a log$_{10}$ scale is used here, rather than the natural logarithm demanded in Eq. \ref{eq:JohnsonSamwer}.}
	\label{fig:SigmaScaling}
\end{figure}
%
Finally we come back to the experimentally extracted stress exponent $n=3/2$ of the stress dependence of activation energy. It is given by the power law $\Delta G _{\mathrm{eff}} = \Delta F_0\, [1- (\tau_a / \hat{\tau})]^{3/2} $ (Eq. \ref{eq:powerlaw}). This barrier height scaling has been shown to be universal for many driven systems, including flowing liquids, mechanically deformed glasses, and stretched proteins \cite{Maloney2006}. Inserting Eq. \ref{eq:powerlaw} into Eq. \ref{eq:strainrate} and setting $T_0 = \Delta F_0 / k_B$, we find
\begin{equation}
\label{eq:JohnsonSamwer}
\tau_a /\hat{\tau} = 1-[\ln (\dot{\gamma}_0/\dot{\gamma}_p)]^{2/3}\,[T/T_0]^{2/3},
\end{equation}
a temperature dependence of stress that agrees with the universal $T^{2/3}$ temperature dependence of plastic yielding in metallic glasses proposed by Johnson and Samwer \cite{Johnson2005,Dasgupta2013}. Verification of this temperature dependence for NC $\mathrm{Pd}_{90}\mathrm{Au}_{10}$ is subject-matter of ongoing work. Nevertheless, it becomes evident that the 
$[\ln(\dot{\gamma}_0/\dot{\gamma}_p)]^{2/3}$-term is not a negligibly small correction term as estimated for metallic glasses at $T < T_g$ where $T_g$ is the glass transition temperature \cite{Johnson2005}. Assuming fixed temperature (RT) and inverting the power law (Eq. \ref{eq:JohnsonSamwer}), it is predicted that $(1-\sigma_\mathrm{a}/\hat\sigma)^{3/2}$ scales as
$\ln \dot{\varepsilon}_p$. In Fig. \ref{fig:SigmaScaling} we display that our experimental data obey the predicted scaling behavior even in the entire microplastic regime. Assuming that each ST is governed by the crossing of a saddle-node bifurcation where the energy barrier assumes the universal scaling form, 
$\Delta G \propto (1 - \tau_a / \hat{\tau})^{3/2}$ , Chattoraj et al. \cite{Chattoraj2010} found similar results in 2D Lenard-Jones glasses. Overall, it seems self-evident to presume that our finding manifests a generalization of the Johnson-Samwer expression $\tau_y /G = a - b (T/T_g)^{2/3}$ where $a,b$ are constants \cite{Johnson2005}. Finally we remark that the observed scaling behavior, covering the entire microplastic regime, suggests that the material response is characteristic of nonlinear viscous behavior, also seen in metallic glasses at temperatures $T>0.6 \, T_g$ \cite{Megusar1979,Johnson2007}. The connotation of strain or work hardening taking place in NC metals in the microplastic regime seems to be misleading then.

\section{Conclusion}
In summary, we conclude that shear transformations \cite{Argon1979} dominate the plastic deformation of NC $\mathrm{Pd}_{90}\mathrm{Au}_{10}$ at the low end of the nanoscale ($\leq \unit[10]{nm}$). Intragranular plasticity based on dislocation glide seems to play a minor role. Likewise, stress-driven grain boundary migration contributes a share of approximately $10 \%$ shear strain to overall strain. Since shear transformations are considered to be the generic flow defect in metallic glasses, it seems reasonable to suppose that the atomic site mismatch (disorder) and the concomitant excess volume in the core regions of grain boundaries provoke the occurrence of shear transformations. We do not stipulate that grain boundaries are structurally -- in terms of atomic short- and mid-range order -- equivalent to metallic glasses. Our conclusion is based on a detailed analysis of thermal activation parameters (activation volume, strain-rate sensitivity, and activation energy) suggesting that grain boundaries can \textit{mimic} shear deformation behavior as observed in metallic glasses without exhibiting stick-slip behavior and catastrophic failure through running away shear bands. By analyzing the stress dependence of activation energy, we find that the energy barrier scales as stress to the power $3/2$. It is straightforward to show that this scaling behavior translates into the universal $T^{2/3}$ temperature dependence of plastic yielding in metallic glasses proposed by Johnson and Samwer \cite{Johnson2005}, and it has been verified for a whole variety of different metallic glasses. Moreover, our analysis reveals that plastic yielding in NC metals with grain sizes of $ \approx \unit[10]{nm}$ or smaller depends markedly on the imposed strain rate $\dot{\varepsilon}$ where stresses $\sigma$ in the entire microplastic regime scale as $\sigma \sim (\ln \dot{\varepsilon})^{2/3}$. From this scaling behavior, we infer that the customary assumption that work or strain hardening takes place in the microplastic regime is not likely to be true. The observed scaling behavior suggests that nonlinear viscous behavior underlies the pronounced increase of stress beyond the regime of linear elasticity up to the yield stress. A similar behavior is observed for metallic glasses when the testing temperature approaches the glass transition temperature from below.         

\section{acknowledgments}
This work has benefited from fruitful discussions with M.J. Demkowicz and A.S. Argon. The authors are grateful for the preparation of the FIB lamellae and TEM dark field micrographs taken by Aaron Kobler at the Karls\-ruhe Nano Micro Facility (KMNF), and for financial support from DFG (FOR 714).

\bibliography{Literatur}

\begin{thebibliography}{65}%
\makeatletter
\providecommand \@ifxundefined [1]{%
 \@ifx{#1\undefined}
}%
\providecommand \@ifnum [1]{%
 \ifnum #1\expandafter \@firstoftwo
 \else \expandafter \@secondoftwo
 \fi
}%
\providecommand \@ifx [1]{%
 \ifx #1\expandafter \@firstoftwo
 \else \expandafter \@secondoftwo
 \fi
}%
\providecommand \natexlab [1]{#1}%
\providecommand \enquote  [1]{``#1''}%
\providecommand \bibnamefont  [1]{#1}%
\providecommand \bibfnamefont [1]{#1}%
\providecommand \citenamefont [1]{#1}%
\providecommand \href@noop [0]{\@secondoftwo}%
\providecommand \href [0]{\begingroup \@sanitize@url \@href}%
\providecommand \@href[1]{\@@startlink{#1}\@@href}%
\providecommand \@@href[1]{\endgroup#1\@@endlink}%
\providecommand \@sanitize@url [0]{\catcode `\\12\catcode `\$12\catcode
  `\&12\catcode `\#12\catcode `\^12\catcode `\_12\catcode `\%12\relax}%
\providecommand \@@startlink[1]{}%
\providecommand \@@endlink[0]{}%
\providecommand \url  [0]{\begingroup\@sanitize@url \@url }%
\providecommand \@url [1]{\endgroup\@href {#1}{\urlprefix }}%
\providecommand \urlprefix  [0]{URL }%
\providecommand \Eprint [0]{\href }%
\providecommand \doibase [0]{http://dx.doi.org/}%
\providecommand \selectlanguage [0]{\@gobble}%
\providecommand \bibinfo  [0]{\@secondoftwo}%
\providecommand \bibfield  [0]{\@secondoftwo}%
\providecommand \translation [1]{[#1]}%
\providecommand \BibitemOpen [0]{}%
\providecommand \bibitemStop [0]{}%
\providecommand \bibitemNoStop [0]{.\EOS\space}%
\providecommand \EOS [0]{\spacefactor3000\relax}%
\providecommand \BibitemShut  [1]{\csname bibitem#1\endcsname}%
\let\auto@bib@innerbib\@empty
\bibitem [{\citenamefont {Argon}(2008)}]{Argon2008}%
  \BibitemOpen
  \bibfield  {author} {\bibinfo {author} {\bibfnamefont {A.~S.}\ \bibnamefont
  {Argon}},\ }\href@noop {} {\emph {\bibinfo {title} {Strengthening Mechanisms
  in Crystal Plasticity}}}\ (\bibinfo  {publisher} {Oxford University Press},\
  \bibinfo {address} {Oxford},\ \bibinfo {year} {2008})\BibitemShut {NoStop}%
\bibitem [{\citenamefont {Argon}(1979)}]{Argon1979}%
  \BibitemOpen
  \bibfield  {author} {\bibinfo {author} {\bibfnamefont {A.~S.}\ \bibnamefont
  {Argon}},\ }\href {\doibase 10.1016/0001-6160(79)90055-5} {\bibfield
  {journal} {\bibinfo  {journal} {Acta Metall.}\ }\textbf {\bibinfo {volume}
  {27}},\ \bibinfo {pages} {47} (\bibinfo {year} {1979})}\BibitemShut {NoStop}%
\bibitem [{\citenamefont {Schuh}\ \emph {et~al.}(2007)\citenamefont {Schuh},
  \citenamefont {Hufnagel},\ and\ \citenamefont {Ramamurty}}]{Schuh2007}%
  \BibitemOpen
  \bibfield  {author} {\bibinfo {author} {\bibfnamefont {C.~A.}\ \bibnamefont
  {Schuh}}, \bibinfo {author} {\bibfnamefont {T.~C.}\ \bibnamefont {Hufnagel}},
  \ and\ \bibinfo {author} {\bibfnamefont {U.}~\bibnamefont {Ramamurty}},\
  }\href {\doibase 10.1016/j.actamat.2007.01.052} {\bibfield  {journal}
  {\bibinfo  {journal} {Acta Mater.}\ }\textbf {\bibinfo {volume} {55}},\
  \bibinfo {pages} {4067} (\bibinfo {year} {2007})}\BibitemShut {NoStop}%
\bibitem [{\citenamefont {Birringer}(1989)}]{Birringer1989}%
  \BibitemOpen
  \bibfield  {author} {\bibinfo {author} {\bibfnamefont {R.}~\bibnamefont
  {Birringer}},\ }\href {\doibase 10.1016/0921-5093(89)90083-X} {\bibfield
  {journal} {\bibinfo  {journal} {Mater. Sci. Eng. A}\ }\textbf {\bibinfo
  {volume} {117}},\ \bibinfo {pages} {33 } (\bibinfo {year}
  {1989})}\BibitemShut {NoStop}%
\bibitem [{\citenamefont {Meyers}\ \emph {et~al.}(2006)\citenamefont {Meyers},
  \citenamefont {Mishra},\ and\ \citenamefont {Benson}}]{Meyers2006}%
  \BibitemOpen
  \bibfield  {author} {\bibinfo {author} {\bibfnamefont {M.~A.}\ \bibnamefont
  {Meyers}}, \bibinfo {author} {\bibfnamefont {A.}~\bibnamefont {Mishra}}, \
  and\ \bibinfo {author} {\bibfnamefont {D.~J.}\ \bibnamefont {Benson}},\
  }\href {\doibase 10.1016/j.pmatsci.2005.08.003} {\bibfield  {journal}
  {\bibinfo  {journal} {Prog. Mater. Sci.}\ }\textbf {\bibinfo {volume} {51}},\
  \bibinfo {pages} {427} (\bibinfo {year} {2006})}\BibitemShut {NoStop}%
\bibitem [{\citenamefont {Dao}\ \emph {et~al.}(2007)\citenamefont {Dao},
  \citenamefont {Lu}, \citenamefont {Asaro}, \citenamefont {De~Hosson},\ and\
  \citenamefont {Ma}}]{Dao2007}%
  \BibitemOpen
  \bibfield  {author} {\bibinfo {author} {\bibfnamefont {M.}~\bibnamefont
  {Dao}}, \bibinfo {author} {\bibfnamefont {L.}~\bibnamefont {Lu}}, \bibinfo
  {author} {\bibfnamefont {R.~J.}\ \bibnamefont {Asaro}}, \bibinfo {author}
  {\bibfnamefont {J.~T.~M.}\ \bibnamefont {De~Hosson}}, \ and\ \bibinfo
  {author} {\bibfnamefont {E.}~\bibnamefont {Ma}},\ }\href {\doibase
  10.1016/j.actamat.2007.01.038} {\bibfield  {journal} {\bibinfo  {journal}
  {Acta Mater.}\ }\textbf {\bibinfo {volume} {55}},\ \bibinfo {pages} {4041 }
  (\bibinfo {year} {2007})}\BibitemShut {NoStop}%
\bibitem [{\citenamefont {Chokshi}\ \emph {et~al.}(1989)\citenamefont
  {Chokshi}, \citenamefont {Rosen}, \citenamefont {Karch},\ and\ \citenamefont
  {Gleiter}}]{Chokshi1989}%
  \BibitemOpen
  \bibfield  {author} {\bibinfo {author} {\bibfnamefont {A.}~\bibnamefont
  {Chokshi}}, \bibinfo {author} {\bibfnamefont {A.}~\bibnamefont {Rosen}},
  \bibinfo {author} {\bibfnamefont {J.}~\bibnamefont {Karch}}, \ and\ \bibinfo
  {author} {\bibfnamefont {H.}~\bibnamefont {Gleiter}},\ }\href {\doibase
  http://dx.doi.org/10.1016/0036-9748(89)90342-6} {\bibfield  {journal}
  {\bibinfo  {journal} {Scripta Metall.}\ }\textbf {\bibinfo {volume} {23}},\
  \bibinfo {pages} {1679} (\bibinfo {year} {1989})}\BibitemShut {NoStop}%
\bibitem [{\citenamefont {Karch}\ \emph {et~al.}(1987)\citenamefont {Karch},
  \citenamefont {Birringer},\ and\ \citenamefont {Gleiter}}]{Karch1987}%
  \BibitemOpen
  \bibfield  {author} {\bibinfo {author} {\bibfnamefont {J.}~\bibnamefont
  {Karch}}, \bibinfo {author} {\bibfnamefont {R.}~\bibnamefont {Birringer}}, \
  and\ \bibinfo {author} {\bibfnamefont {H.}~\bibnamefont {Gleiter}},\ }\href
  {\doibase 10.1038/330556a0} {\bibfield  {journal} {\bibinfo  {journal}
  {Nature}\ }\textbf {\bibinfo {volume} {330}},\ \bibinfo {pages} {556 }
  (\bibinfo {year} {1987})}\BibitemShut {NoStop}%
\bibitem [{\citenamefont {Vo}\ \emph {et~al.}(2008)\citenamefont {Vo},
  \citenamefont {Averback}, \citenamefont {Bellon},\ and\ \citenamefont
  {Caro}}]{Vo2008}%
  \BibitemOpen
  \bibfield  {author} {\bibinfo {author} {\bibfnamefont {N.~Q.}\ \bibnamefont
  {Vo}}, \bibinfo {author} {\bibfnamefont {R.~S.}\ \bibnamefont {Averback}},
  \bibinfo {author} {\bibfnamefont {P.}~\bibnamefont {Bellon}}, \ and\ \bibinfo
  {author} {\bibfnamefont {A.}~\bibnamefont {Caro}},\ }\href {\doibase
  10.1103/PhysRevB.78.241402} {\bibfield  {journal} {\bibinfo  {journal} {Phys.
  Rev. B}\ }\textbf {\bibinfo {volume} {78}},\ \bibinfo {pages} {241402}
  (\bibinfo {year} {2008})}\BibitemShut {NoStop}%
\bibitem [{\citenamefont {Van~Swygenhoven}\ and\ \citenamefont
  {Derlet}(2001)}]{VanSwygenhoven2001}%
  \BibitemOpen
  \bibfield  {author} {\bibinfo {author} {\bibfnamefont {H.}~\bibnamefont
  {Van~Swygenhoven}}\ and\ \bibinfo {author} {\bibfnamefont {P.~M.}\
  \bibnamefont {Derlet}},\ }\href {\doibase 10.1103/PhysRevB.64.224105}
  {\bibfield  {journal} {\bibinfo  {journal} {Phys. Rev. B}\ }\textbf {\bibinfo
  {volume} {64}},\ \bibinfo {pages} {224105} (\bibinfo {year}
  {2001})}\BibitemShut {NoStop}%
\bibitem [{\citenamefont {Weissm\"{u}ller}\ \emph {et~al.}(2011)\citenamefont
  {Weissm\"{u}ller}, \citenamefont {Markmann}, \citenamefont {Grewer},\ and\
  \citenamefont {Birringer}}]{Weissmuller2011}%
  \BibitemOpen
  \bibfield  {author} {\bibinfo {author} {\bibfnamefont {J.}~\bibnamefont
  {Weissm\"{u}ller}}, \bibinfo {author} {\bibfnamefont {J.}~\bibnamefont
  {Markmann}}, \bibinfo {author} {\bibfnamefont {M.}~\bibnamefont {Grewer}}, \
  and\ \bibinfo {author} {\bibfnamefont {R.}~\bibnamefont {Birringer}},\ }\href
  {\doibase 10.1016/j.actamat.2011.03.060} {\bibfield  {journal} {\bibinfo
  {journal} {Acta Mater.}\ }\textbf {\bibinfo {volume} {59}},\ \bibinfo {pages}
  {4366} (\bibinfo {year} {2011})}\BibitemShut {NoStop}%
\bibitem [{\citenamefont {Cahn}\ \emph {et~al.}(2006)\citenamefont {Cahn},
  \citenamefont {Mishin},\ and\ \citenamefont {Suzuki}}]{Cahn2006}%
  \BibitemOpen
  \bibfield  {author} {\bibinfo {author} {\bibfnamefont {J.~W.}\ \bibnamefont
  {Cahn}}, \bibinfo {author} {\bibfnamefont {Y.}~\bibnamefont {Mishin}}, \ and\
  \bibinfo {author} {\bibfnamefont {A.}~\bibnamefont {Suzuki}},\ }\href
  {\doibase 10.1016/j.actamat.2006.08.004} {\bibfield  {journal} {\bibinfo
  {journal} {Acta Mater.}\ }\textbf {\bibinfo {volume} {54}},\ \bibinfo {pages}
  {4953} (\bibinfo {year} {2006})}\BibitemShut {NoStop}%
\bibitem [{\citenamefont {Cahn}\ and\ \citenamefont {Taylor}(2004)}]{Cahn2004}%
  \BibitemOpen
  \bibfield  {author} {\bibinfo {author} {\bibfnamefont {J.~W.}\ \bibnamefont
  {Cahn}}\ and\ \bibinfo {author} {\bibfnamefont {J.~E.}\ \bibnamefont
  {Taylor}},\ }\href {\doibase 10.1016/j.actamat.2004.02.048} {\bibfield
  {journal} {\bibinfo  {journal} {Acta Mater.}\ }\textbf {\bibinfo {volume}
  {52}},\ \bibinfo {pages} {4887} (\bibinfo {year} {2004})}\BibitemShut
  {NoStop}%
\bibitem [{\citenamefont {Legros}\ \emph {et~al.}(2008)\citenamefont {Legros},
  \citenamefont {Gianola},\ and\ \citenamefont {Hemker}}]{Legros2008}%
  \BibitemOpen
  \bibfield  {author} {\bibinfo {author} {\bibfnamefont {M.}~\bibnamefont
  {Legros}}, \bibinfo {author} {\bibfnamefont {D.~S.}\ \bibnamefont {Gianola}},
  \ and\ \bibinfo {author} {\bibfnamefont {K.~J.}\ \bibnamefont {Hemker}},\
  }\href {\doibase 10.1016/j.actamat.2008.03.032} {\bibfield  {journal}
  {\bibinfo  {journal} {Acta Mater.}\ }\textbf {\bibinfo {volume} {56}},\
  \bibinfo {pages} {3380 } (\bibinfo {year} {2008})}\BibitemShut {NoStop}%
\bibitem [{\citenamefont {Lund}\ and\ \citenamefont {Schuh}(2005)}]{Lund2005}%
  \BibitemOpen
  \bibfield  {author} {\bibinfo {author} {\bibfnamefont {A.~C.}\ \bibnamefont
  {Lund}}\ and\ \bibinfo {author} {\bibfnamefont {C.~A.}\ \bibnamefont
  {Schuh}},\ }\href {\doibase 10.1016/j.actamat.2005.03.023} {\bibfield
  {journal} {\bibinfo  {journal} {Acta Mater.}\ }\textbf {\bibinfo {volume}
  {53}},\ \bibinfo {pages} {3193} (\bibinfo {year} {2005})}\BibitemShut
  {NoStop}%
\bibitem [{\citenamefont {Argon}\ and\ \citenamefont {Yip}(2006)}]{Argon2006}%
  \BibitemOpen
  \bibfield  {author} {\bibinfo {author} {\bibfnamefont {A.~S.}\ \bibnamefont
  {Argon}}\ and\ \bibinfo {author} {\bibfnamefont {S.}~\bibnamefont {Yip}},\
  }\href {\doibase 10.1080/09500830600986091} {\bibfield  {journal} {\bibinfo
  {journal} {Philos. Mag. Lett.}\ }\textbf {\bibinfo {volume} {86}},\ \bibinfo
  {pages} {713} (\bibinfo {year} {2006})}\BibitemShut {NoStop}%
\bibitem [{\citenamefont {Gu}\ \emph {et~al.}(2011)\citenamefont {Gu},
  \citenamefont {Dao}, \citenamefont {Asaro},\ and\ \citenamefont
  {Suresh}}]{Gu2011}%
  \BibitemOpen
  \bibfield  {author} {\bibinfo {author} {\bibfnamefont {P.}~\bibnamefont
  {Gu}}, \bibinfo {author} {\bibfnamefont {M.}~\bibnamefont {Dao}}, \bibinfo
  {author} {\bibfnamefont {R.~J.}\ \bibnamefont {Asaro}}, \ and\ \bibinfo
  {author} {\bibfnamefont {S.}~\bibnamefont {Suresh}},\ }\href {\doibase
  10.1016/j.actamat.2011.07.019} {\bibfield  {journal} {\bibinfo  {journal}
  {Acta Mater.}\ }\textbf {\bibinfo {volume} {59}},\ \bibinfo {pages} {6861 }
  (\bibinfo {year} {2011})}\BibitemShut {NoStop}%
\bibitem [{\citenamefont {Asaro}\ and\ \citenamefont
  {Suresh}(2005)}]{Asaro2005}%
  \BibitemOpen
  \bibfield  {author} {\bibinfo {author} {\bibfnamefont {R.~J.}\ \bibnamefont
  {Asaro}}\ and\ \bibinfo {author} {\bibfnamefont {S.}~\bibnamefont {Suresh}},\
  }\href {\doibase 10.1016/j.actamat.2005.03.047} {\bibfield  {journal}
  {\bibinfo  {journal} {Acta Mater.}\ }\textbf {\bibinfo {volume} {53}},\
  \bibinfo {pages} {3369 } (\bibinfo {year} {2005})}\BibitemShut {NoStop}%
\bibitem [{\citenamefont {Ames}\ \emph {et~al.}(2012)\citenamefont {Ames},
  \citenamefont {Grewer}, \citenamefont {Braun},\ and\ \citenamefont
  {Birringer}}]{Ames2012}%
  \BibitemOpen
  \bibfield  {author} {\bibinfo {author} {\bibfnamefont {M.}~\bibnamefont
  {Ames}}, \bibinfo {author} {\bibfnamefont {M.}~\bibnamefont {Grewer}},
  \bibinfo {author} {\bibfnamefont {C.}~\bibnamefont {Braun}}, \ and\ \bibinfo
  {author} {\bibfnamefont {R.}~\bibnamefont {Birringer}},\ }\href {\doibase
  10.1016/j.msea.2012.03.061} {\bibfield  {journal} {\bibinfo  {journal}
  {Mater. Sci. Eng., A}\ }\textbf {\bibinfo {volume} {546}},\ \bibinfo {pages}
  {248} (\bibinfo {year} {2012})}\BibitemShut {NoStop}%
\bibitem [{\citenamefont {Seeger}(1955)}]{Seeger1955}%
  \BibitemOpen
  \bibfield  {author} {\bibinfo {author} {\bibfnamefont {A.}~\bibnamefont
  {Seeger}},\ }\href@noop {} {\emph {\bibinfo {title} {Encyclopedia of Physics
  Vol. VII Part 1.}}},\ edited by\ \bibinfo {editor} {\bibfnamefont
  {S.}~\bibnamefont {Fl\"{u}gge}}\ (\bibinfo  {publisher} {Springer-Verlag},\
  \bibinfo {address} {Berlin},\ \bibinfo {year} {1955})\BibitemShut {NoStop}%
\bibitem [{\citenamefont {Krill}\ and\ \citenamefont
  {Birringer}(1998)}]{Krill1998}%
  \BibitemOpen
  \bibfield  {author} {\bibinfo {author} {\bibfnamefont {C.~E.}\ \bibnamefont
  {Krill}}\ and\ \bibinfo {author} {\bibfnamefont {R.}~\bibnamefont
  {Birringer}},\ }\href@noop {} {\bibfield  {journal} {\bibinfo  {journal}
  {Phil. Mag. A}\ }\textbf {\bibinfo {volume} {77}},\ \bibinfo {pages} {621}
  (\bibinfo {year} {1998})}\BibitemShut {NoStop}%
\bibitem [{\citenamefont {Markmann}\ \emph {et~al.}(2003)\citenamefont
  {Markmann}, \citenamefont {Bunzel}, \citenamefont {R\"{o}sner}, \citenamefont
  {Liu}, \citenamefont {Padmanabhan}, \citenamefont {Birringer}, \citenamefont
  {Gleiter},\ and\ \citenamefont {Weissm\"{u}ller}}]{Markmann2003}%
  \BibitemOpen
  \bibfield  {author} {\bibinfo {author} {\bibfnamefont {J.}~\bibnamefont
  {Markmann}}, \bibinfo {author} {\bibfnamefont {P.}~\bibnamefont {Bunzel}},
  \bibinfo {author} {\bibfnamefont {H.}~\bibnamefont {R\"{o}sner}}, \bibinfo
  {author} {\bibfnamefont {K.~W.}\ \bibnamefont {Liu}}, \bibinfo {author}
  {\bibfnamefont {K.~A.}\ \bibnamefont {Padmanabhan}}, \bibinfo {author}
  {\bibfnamefont {R.}~\bibnamefont {Birringer}}, \bibinfo {author}
  {\bibfnamefont {H.}~\bibnamefont {Gleiter}}, \ and\ \bibinfo {author}
  {\bibfnamefont {J.}~\bibnamefont {Weissm\"{u}ller}},\ }\href {\doibase
  10.1016/S1359-6462(03)00401-9} {\bibfield  {journal} {\bibinfo  {journal}
  {Scr. Mater.}\ }\textbf {\bibinfo {volume} {49}},\ \bibinfo {pages} {637}
  (\bibinfo {year} {2003})}\BibitemShut {NoStop}%
\bibitem [{\citenamefont {Schaefer}\ \emph {et~al.}(2000)\citenamefont
  {Schaefer}, \citenamefont {Reimann}, \citenamefont {Straub}, \citenamefont
  {Phillipp}, \citenamefont {Tanimoto}, \citenamefont {Brossmann},\ and\
  \citenamefont {Würschum}}]{Schaefer2000}%
  \BibitemOpen
  \bibfield  {author} {\bibinfo {author} {\bibfnamefont {H.-E.}\ \bibnamefont
  {Schaefer}}, \bibinfo {author} {\bibfnamefont {K.}~\bibnamefont {Reimann}},
  \bibinfo {author} {\bibfnamefont {W.}~\bibnamefont {Straub}}, \bibinfo
  {author} {\bibfnamefont {F.}~\bibnamefont {Phillipp}}, \bibinfo {author}
  {\bibfnamefont {H.}~\bibnamefont {Tanimoto}}, \bibinfo {author}
  {\bibfnamefont {U.}~\bibnamefont {Brossmann}}, \ and\ \bibinfo {author}
  {\bibfnamefont {R.}~\bibnamefont {Würschum}},\ }\href {\doibase
  http://dx.doi.org/10.1016/S0921-5093(00)00659-6} {\bibfield  {journal}
  {\bibinfo  {journal} {Mater. Sci. Eng. A}\ }\textbf {\bibinfo {volume}
  {286}},\ \bibinfo {pages} {24} (\bibinfo {year} {2000})}\BibitemShut
  {NoStop}%
\bibitem [{\citenamefont {Birringer}\ \emph {et~al.}(2002)\citenamefont
  {Birringer}, \citenamefont {Hoffmann},\ and\ \citenamefont
  {Zimmer}}]{Birringer2002}%
  \BibitemOpen
  \bibfield  {author} {\bibinfo {author} {\bibfnamefont {R.}~\bibnamefont
  {Birringer}}, \bibinfo {author} {\bibfnamefont {M.}~\bibnamefont {Hoffmann}},
  \ and\ \bibinfo {author} {\bibfnamefont {P.}~\bibnamefont {Zimmer}},\
  }\href@noop {} {\bibfield  {journal} {\bibinfo  {journal} {Phys. Rev. Lett.}\
  }\textbf {\bibinfo {volume} {88}},\ \bibinfo {pages} {206104(4)} (\bibinfo
  {year} {2002})}\BibitemShut {NoStop}%
\bibitem [{\citenamefont {Rittel}\ \emph {et~al.}(2002)\citenamefont {Rittel},
  \citenamefont {Lee},\ and\ \citenamefont {Ravichandran}}]{Rittel2002}%
  \BibitemOpen
  \bibfield  {author} {\bibinfo {author} {\bibfnamefont {D.}~\bibnamefont
  {Rittel}}, \bibinfo {author} {\bibfnamefont {S.}~\bibnamefont {Lee}}, \ and\
  \bibinfo {author} {\bibfnamefont {G.}~\bibnamefont {Ravichandran}},\ }\href
  {\doibase 10.1007/BF02411052} {\bibfield  {journal} {\bibinfo  {journal}
  {Exp. Mech.}\ }\textbf {\bibinfo {volume} {42}},\ \bibinfo {pages} {58}
  (\bibinfo {year} {2002})}\BibitemShut {NoStop}%
\bibitem [{\citenamefont {Ames}\ \emph {et~al.}(2010)\citenamefont {Ames},
  \citenamefont {Markmann},\ and\ \citenamefont {Birringer}}]{Ames2010}%
  \BibitemOpen
  \bibfield  {author} {\bibinfo {author} {\bibfnamefont {M.}~\bibnamefont
  {Ames}}, \bibinfo {author} {\bibfnamefont {J.}~\bibnamefont {Markmann}}, \
  and\ \bibinfo {author} {\bibfnamefont {R.}~\bibnamefont {Birringer}},\ }\href
  {\doibase 10.1016/j.msea.2010.09.049} {\bibfield  {journal} {\bibinfo
  {journal} {Mater. Sci. Eng., A}\ }\textbf {\bibinfo {volume} {528}},\
  \bibinfo {pages} {526} (\bibinfo {year} {2010})}\BibitemShut {NoStop}%
\bibitem [{\citenamefont {Grewer}\ \emph {et~al.}(2011)\citenamefont {Grewer},
  \citenamefont {Markmann}, \citenamefont {Karos}, \citenamefont {Arnold},\
  and\ \citenamefont {Birringer}}]{Grewer2011}%
  \BibitemOpen
  \bibfield  {author} {\bibinfo {author} {\bibfnamefont {M.}~\bibnamefont
  {Grewer}}, \bibinfo {author} {\bibfnamefont {J.}~\bibnamefont {Markmann}},
  \bibinfo {author} {\bibfnamefont {R.}~\bibnamefont {Karos}}, \bibinfo
  {author} {\bibfnamefont {W.}~\bibnamefont {Arnold}}, \ and\ \bibinfo {author}
  {\bibfnamefont {R.}~\bibnamefont {Birringer}},\ }\href {\doibase
  10.1016/j.actamat.2010.11.016} {\bibfield  {journal} {\bibinfo  {journal}
  {Acta Mater.}\ }\textbf {\bibinfo {volume} {59}},\ \bibinfo {pages} {1523 }
  (\bibinfo {year} {2011})}\BibitemShut {NoStop}%
\bibitem [{\citenamefont {Saada}\ and\ \citenamefont
  {Kruml}(2011)}]{Saada2011}%
  \BibitemOpen
  \bibfield  {author} {\bibinfo {author} {\bibfnamefont {G.}~\bibnamefont
  {Saada}}\ and\ \bibinfo {author} {\bibfnamefont {T.}~\bibnamefont {Kruml}},\
  }\href {\doibase 10.1016/j.actamat.2010.12.035} {\bibfield  {journal}
  {\bibinfo  {journal} {Acta Mater.}\ }\textbf {\bibinfo {volume} {59}},\
  \bibinfo {pages} {2565 } (\bibinfo {year} {2011})}\BibitemShut {NoStop}%
\bibitem [{Note1()}]{Note1}%
  \BibitemOpen
  \bibinfo {note} {We assumed a value of $\protect \unit [0.275]{nm}$ for the
  magnitude of the burgers vector $\protect \mathrm {b}$}\BibitemShut {NoStop}%
\bibitem [{\citenamefont {Frost}\ and\ \citenamefont
  {Ashby}(1982)}]{Frost1982}%
  \BibitemOpen
  \bibfield  {author} {\bibinfo {author} {\bibfnamefont {H.~J.}\ \bibnamefont
  {Frost}}\ and\ \bibinfo {author} {\bibfnamefont {M.~F.}\ \bibnamefont
  {Ashby}},\ }\href@noop {} {\emph {\bibinfo {title} {Deformation-Mechanism
  Maps: The Plasticity and Creep of Metals and Ceramics}}}\ (\bibinfo
  {publisher} {Pergamon Press},\ \bibinfo {address} {Oxford},\ \bibinfo {year}
  {1982})\ p.\ \bibinfo {pages} {165}\BibitemShut {NoStop}%
\bibitem [{\citenamefont {Herring}(1950)}]{Herring1950}%
  \BibitemOpen
  \bibfield  {author} {\bibinfo {author} {\bibfnamefont {C.}~\bibnamefont
  {Herring}},\ }\href {\doibase http://dx.doi.org/10.1063/1.1699681} {\bibfield
   {journal} {\bibinfo  {journal} {J. Appl. Phys.}\ }\textbf {\bibinfo {volume}
  {21}},\ \bibinfo {pages} {437} (\bibinfo {year} {1950})}\BibitemShut
  {NoStop}%
\bibitem [{\citenamefont {Coble}(1963)}]{Coble1963}%
  \BibitemOpen
  \bibfield  {author} {\bibinfo {author} {\bibfnamefont {R.~L.}\ \bibnamefont
  {Coble}},\ }\href {\doibase DOI:10.1063/1.1702656} {\bibfield  {journal}
  {\bibinfo  {journal} {J. Appl. Phys.}\ }\textbf {\bibinfo {volume} {34}},\
  \bibinfo {pages} {1679} (\bibinfo {year} {1963})}\BibitemShut {NoStop}%
\bibitem [{\citenamefont {Langdon}(2006)}]{Langdon2006}%
  \BibitemOpen
  \bibfield  {author} {\bibinfo {author} {\bibfnamefont {T.~G.}\ \bibnamefont
  {Langdon}},\ }\href {\doibase 10.1007/s10853-006-6476-0} {\bibfield
  {journal} {\bibinfo  {journal} {J. Mater. Sci.}\ }\textbf {\bibinfo {volume}
  {41}},\ \bibinfo {pages} {597} (\bibinfo {year} {2006})}\BibitemShut
  {NoStop}%
\bibitem [{\citenamefont {L\"{u}thy}\ \emph {et~al.}(1979)\citenamefont
  {L\"{u}thy}, \citenamefont {White},\ and\ \citenamefont
  {Sherby}}]{Luthy1979}%
  \BibitemOpen
  \bibfield  {author} {\bibinfo {author} {\bibfnamefont {H.}~\bibnamefont
  {L\"{u}thy}}, \bibinfo {author} {\bibfnamefont {R.~A.}\ \bibnamefont
  {White}}, \ and\ \bibinfo {author} {\bibfnamefont {O.~D.}\ \bibnamefont
  {Sherby}},\ }\href {\doibase 10.1016/0025-5416(79)90060-0} {\bibfield
  {journal} {\bibinfo  {journal} {Mater. Sci. Eng.}\ }\textbf {\bibinfo
  {volume} {39}},\ \bibinfo {pages} {211 } (\bibinfo {year}
  {1979})}\BibitemShut {NoStop}%
\bibitem [{\citenamefont {Rachinger}(1952)}]{Rachinger1952}%
  \BibitemOpen
  \bibfield  {author} {\bibinfo {author} {\bibfnamefont {W.}~\bibnamefont
  {Rachinger}},\ }\href@noop {} {\bibfield  {journal} {\bibinfo  {journal} {J.
  Inst. Met.}\ }\textbf {\bibinfo {volume} {81}},\ \bibinfo {pages} {33}
  (\bibinfo {year} {1952})}\BibitemShut {NoStop}%
\bibitem [{\citenamefont {Lifshitz}(1963)}]{Lifshitz1963a}%
  \BibitemOpen
  \bibfield  {author} {\bibinfo {author} {\bibfnamefont {I.~M.}\ \bibnamefont
  {Lifshitz}},\ }\href@noop {} {\bibfield  {journal} {\bibinfo  {journal}
  {Soviet. Phys. JETP}\ }\textbf {\bibinfo {volume} {17}},\ \bibinfo {pages}
  {909} (\bibinfo {year} {1963})}\BibitemShut {NoStop}%
\bibitem [{\citenamefont {Argon}\ and\ \citenamefont
  {Demkowicz}(2006)}]{Argon2006a}%
  \BibitemOpen
  \bibfield  {author} {\bibinfo {author} {\bibfnamefont {A.~S.}\ \bibnamefont
  {Argon}}\ and\ \bibinfo {author} {\bibfnamefont {M.~J.}\ \bibnamefont
  {Demkowicz}},\ }\href {\doibase 10.1080/14786430600596852} {\bibfield
  {journal} {\bibinfo  {journal} {Philos. Mag.}\ }\textbf {\bibinfo {volume}
  {86}},\ \bibinfo {pages} {4153} (\bibinfo {year} {2006})}\BibitemShut
  {NoStop}%
\bibitem [{\citenamefont {Demkowicz}\ and\ \citenamefont
  {Argon}(2005)}]{Demkowicz2005}%
  \BibitemOpen
  \bibfield  {author} {\bibinfo {author} {\bibfnamefont {M.~J.}\ \bibnamefont
  {Demkowicz}}\ and\ \bibinfo {author} {\bibfnamefont {A.~S.}\ \bibnamefont
  {Argon}},\ }\href {\doibase 10.1103/PhysRevB.72.245205} {\bibfield  {journal}
  {\bibinfo  {journal} {Phys. Rev. B}\ }\textbf {\bibinfo {volume} {72}},\
  \bibinfo {pages} {245205} (\bibinfo {year} {2005})}\BibitemShut {NoStop}%
\bibitem [{\citenamefont {Dahmen}\ \emph {et~al.}(2009)\citenamefont {Dahmen},
  \citenamefont {Ben-Zion},\ and\ \citenamefont {Uhl}}]{Dahmen2009}%
  \BibitemOpen
  \bibfield  {author} {\bibinfo {author} {\bibfnamefont {K.~A.}\ \bibnamefont
  {Dahmen}}, \bibinfo {author} {\bibfnamefont {Y.}~\bibnamefont {Ben-Zion}}, \
  and\ \bibinfo {author} {\bibfnamefont {J.~T.}\ \bibnamefont {Uhl}},\ }\href
  {\doibase 10.1103/PhysRevLett.102.175501} {\bibfield  {journal} {\bibinfo
  {journal} {Phys. Rev. Lett.}\ }\textbf {\bibinfo {volume} {102}},\ \bibinfo
  {pages} {175501} (\bibinfo {year} {2009})}\BibitemShut {NoStop}%
\bibitem [{\citenamefont {Maloney}\ and\ \citenamefont
  {Lacks}(2006)}]{Maloney2006}%
  \BibitemOpen
  \bibfield  {author} {\bibinfo {author} {\bibfnamefont {C.~E.}\ \bibnamefont
  {Maloney}}\ and\ \bibinfo {author} {\bibfnamefont {D.~J.}\ \bibnamefont
  {Lacks}},\ }\href {\doibase 10.1103/PhysRevE.73.061106} {\bibfield  {journal}
  {\bibinfo  {journal} {Phys. Rev. E}\ }\textbf {\bibinfo {volume} {73}},\
  \bibinfo {pages} {061106} (\bibinfo {year} {2006})}\BibitemShut {NoStop}%
\bibitem [{\citenamefont {Argon}(2013)}]{Argon2013}%
  \BibitemOpen
  \bibfield  {author} {\bibinfo {author} {\bibfnamefont {A.~S.}\ \bibnamefont
  {Argon}},\ }\href@noop {} {\emph {\bibinfo {title} {The physics of
  deformation and fracture of polymers}}}\ (\bibinfo  {publisher} {Cambridge
  University Press},\ \bibinfo {address} {Cambridge},\ \bibinfo {year}
  {2013})\BibitemShut {NoStop}%
\bibitem [{\citenamefont {Jensen}(2000)}]{Jensen2000}%
  \BibitemOpen
  \bibfield  {author} {\bibinfo {author} {\bibfnamefont {H.~J.}\ \bibnamefont
  {Jensen}},\ }\href@noop {} {\emph {\bibinfo {title} {Self-organized
  criticality : emergent complex behavior in physical and biological
  systems}}}\ (\bibinfo  {publisher} {Cambridge University Press},\ \bibinfo
  {address} {Cambridge},\ \bibinfo {year} {2000})\BibitemShut {NoStop}%
\bibitem [{\citenamefont {Molodov}\ \emph {et~al.}(2011)\citenamefont
  {Molodov}, \citenamefont {Gorkaya},\ and\ \citenamefont
  {Gottstein}}]{Molodov2011}%
  \BibitemOpen
  \bibfield  {author} {\bibinfo {author} {\bibfnamefont {D.}~\bibnamefont
  {Molodov}}, \bibinfo {author} {\bibfnamefont {T.}~\bibnamefont {Gorkaya}}, \
  and\ \bibinfo {author} {\bibfnamefont {G.}~\bibnamefont {Gottstein}},\ }\href
  {\doibase 10.1007/s10853-010-5233-6} {\bibfield  {journal} {\bibinfo
  {journal} {J. Mater. Sci.}\ }\textbf {\bibinfo {volume} {46}},\ \bibinfo
  {pages} {4318} (\bibinfo {year} {2011})}\BibitemShut {NoStop}%
\bibitem [{\citenamefont {Pan}\ \emph {et~al.}(2008)\citenamefont {Pan},
  \citenamefont {Inoue}, \citenamefont {Sakurai},\ and\ \citenamefont
  {Chen}}]{Pan2008}%
  \BibitemOpen
  \bibfield  {author} {\bibinfo {author} {\bibfnamefont {D.}~\bibnamefont
  {Pan}}, \bibinfo {author} {\bibfnamefont {A.}~\bibnamefont {Inoue}}, \bibinfo
  {author} {\bibfnamefont {T.}~\bibnamefont {Sakurai}}, \ and\ \bibinfo
  {author} {\bibfnamefont {M.~W.}\ \bibnamefont {Chen}},\ }\href {\doibase
  10.1073 / pnas.0806051105} {\bibfield  {journal} {\bibinfo  {journal} {PNAS}\
  }\textbf {\bibinfo {volume} {105}},\ \bibinfo {pages} {14769} (\bibinfo
  {year} {2008})}\BibitemShut {NoStop}%
\bibitem [{\citenamefont {Ju}\ \emph {et~al.}(2011)\citenamefont {Ju},
  \citenamefont {Jang}, \citenamefont {Nwankpa},\ and\ \citenamefont
  {Atzmon}}]{Ju2011}%
  \BibitemOpen
  \bibfield  {author} {\bibinfo {author} {\bibfnamefont {J.~D.}\ \bibnamefont
  {Ju}}, \bibinfo {author} {\bibfnamefont {D.}~\bibnamefont {Jang}}, \bibinfo
  {author} {\bibfnamefont {A.}~\bibnamefont {Nwankpa}}, \ and\ \bibinfo
  {author} {\bibfnamefont {M.}~\bibnamefont {Atzmon}},\ }\href {\doibase
  10.1063/1.3552300} {\bibfield  {journal} {\bibinfo  {journal} {J. Appl.
  Phys.}\ }\textbf {\bibinfo {volume} {109}},\ \bibinfo {eid} {053522}
  (\bibinfo {year} {2011})}\BibitemShut {NoStop}%
\bibitem [{\citenamefont {Skrotzki}\ \emph {et~al.}(2013)\citenamefont
  {Skrotzki}, \citenamefont {Eschke}, \citenamefont {J\'oni}, \citenamefont
  {Ung\'ar}, \citenamefont {T\'oth}, \citenamefont {Ivanisenko},\ and\
  \citenamefont {Kurmanaeva}}]{Skrotzki2013}%
  \BibitemOpen
  \bibfield  {author} {\bibinfo {author} {\bibfnamefont {W.}~\bibnamefont
  {Skrotzki}}, \bibinfo {author} {\bibfnamefont {A.}~\bibnamefont {Eschke}},
  \bibinfo {author} {\bibfnamefont {B.}~\bibnamefont {J\'oni}}, \bibinfo
  {author} {\bibfnamefont {T.}~\bibnamefont {Ung\'ar}}, \bibinfo {author}
  {\bibfnamefont {L.}~\bibnamefont {T\'oth}}, \bibinfo {author} {\bibfnamefont
  {Y.}~\bibnamefont {Ivanisenko}}, \ and\ \bibinfo {author} {\bibfnamefont
  {L.}~\bibnamefont {Kurmanaeva}},\ }\href {\doibase
  http://dx.doi.org/10.1016/j.actamat.2013.08.032} {\bibfield  {journal}
  {\bibinfo  {journal} {Acta Mater.}\ }\textbf {\bibinfo {volume} {61}},\
  \bibinfo {pages} {7271 } (\bibinfo {year} {2013})}\BibitemShut {NoStop}%
\bibitem [{\citenamefont {Birringer}\ \emph {et~al.}(1995)\citenamefont
  {Birringer}, \citenamefont {Krill},\ and\ \citenamefont
  {Klingel}}]{Birringer1995}%
  \BibitemOpen
  \bibfield  {author} {\bibinfo {author} {\bibfnamefont {R.}~\bibnamefont
  {Birringer}}, \bibinfo {author} {\bibfnamefont {C.~E.}\ \bibnamefont
  {Krill}}, \ and\ \bibinfo {author} {\bibfnamefont {M.}~\bibnamefont
  {Klingel}},\ }\href@noop {} {\bibfield  {journal} {\bibinfo  {journal} {Phil.
  Mag. Lett.}\ }\textbf {\bibinfo {volume} {72}},\ \bibinfo {pages} {71}
  (\bibinfo {year} {1995})}\BibitemShut {NoStop}%
\bibitem [{\citenamefont {Krill}\ \emph {et~al.}(2001)\citenamefont {Krill},
  \citenamefont {Helfen}, \citenamefont {Michels}, \citenamefont {Natter},
  \citenamefont {Fitch}, \citenamefont {Masson},\ and\ \citenamefont
  {Birringer}}]{Krill2001}%
  \BibitemOpen
  \bibfield  {author} {\bibinfo {author} {\bibfnamefont {C.~E.}\ \bibnamefont
  {Krill}}, \bibinfo {author} {\bibfnamefont {L.}~\bibnamefont {Helfen}},
  \bibinfo {author} {\bibfnamefont {D.}~\bibnamefont {Michels}}, \bibinfo
  {author} {\bibfnamefont {H.}~\bibnamefont {Natter}}, \bibinfo {author}
  {\bibfnamefont {A.}~\bibnamefont {Fitch}}, \bibinfo {author} {\bibfnamefont
  {O.}~\bibnamefont {Masson}}, \ and\ \bibinfo {author} {\bibfnamefont
  {R.}~\bibnamefont {Birringer}},\ }\href {\doibase 10.1103/PhysRevLett.86.842}
  {\bibfield  {journal} {\bibinfo  {journal} {Phys. Rev. Lett.}\ }\textbf
  {\bibinfo {volume} {86}},\ \bibinfo {pages} {842} (\bibinfo {year}
  {2001})}\BibitemShut {NoStop}%
\bibitem [{\citenamefont {Cahn}\ and\ \citenamefont
  {Nabarro}(2001)}]{Cahn2001}%
  \BibitemOpen
  \bibfield  {author} {\bibinfo {author} {\bibfnamefont {J.~W.}\ \bibnamefont
  {Cahn}}\ and\ \bibinfo {author} {\bibfnamefont {F.~N.}\ \bibnamefont
  {Nabarro}},\ }\href {\doibase 10.1080/01418610108214448} {\bibfield
  {journal} {\bibinfo  {journal} {Phil. Mag. A}\ }\textbf {\bibinfo {volume}
  {81}},\ \bibinfo {pages} {1409} (\bibinfo {year} {2001})}\BibitemShut
  {NoStop}%
\bibitem [{\citenamefont {Cottrell}(2002)}]{Cottrell2002}%
  \BibitemOpen
  \bibfield  {author} {\bibinfo {author} {\bibfnamefont {A.~H.}\ \bibnamefont
  {Cottrell}},\ }\href {\doibase 10.1080/09500830110104297} {\bibfield
  {journal} {\bibinfo  {journal} {Philos. Mag. Lett.}\ }\textbf {\bibinfo
  {volume} {82}},\ \bibinfo {pages} {65} (\bibinfo {year} {2002})}\BibitemShut
  {NoStop}%
\bibitem [{\citenamefont {Bulatov}\ and\ \citenamefont
  {Argon}(1994)}]{Bulatov1994b}%
  \BibitemOpen
  \bibfield  {author} {\bibinfo {author} {\bibfnamefont {V.~V.}\ \bibnamefont
  {Bulatov}}\ and\ \bibinfo {author} {\bibfnamefont {A.~S.}\ \bibnamefont
  {Argon}},\ }\href {\doibase 10.1088/0965-0393/2/2/003} {\bibfield  {journal}
  {\bibinfo  {journal} {Modelling Simul. Mater. Sci. Eng.}\ }\textbf {\bibinfo
  {volume} {2}},\ \bibinfo {pages} {203} (\bibinfo {year} {1994})}\BibitemShut
  {NoStop}%
\bibitem [{\citenamefont {Eshelby}(1957)}]{Eshelby1957}%
  \BibitemOpen
  \bibfield  {author} {\bibinfo {author} {\bibfnamefont {J.~D.}\ \bibnamefont
  {Eshelby}},\ }\href {\doibase 10.1098/rspa.1957.0133} {\bibfield  {journal}
  {\bibinfo  {journal} {Proc. R. Soc. Lond. A}\ }\textbf {\bibinfo {volume}
  {241}},\ \bibinfo {pages} {376} (\bibinfo {year} {1957})}\BibitemShut
  {NoStop}%
\bibitem [{\citenamefont {Kocks}\ and\ \citenamefont
  {Mecking}(2003)}]{Kocks2003}%
  \BibitemOpen
  \bibfield  {author} {\bibinfo {author} {\bibfnamefont {U.~F.}\ \bibnamefont
  {Kocks}}\ and\ \bibinfo {author} {\bibfnamefont {H.}~\bibnamefont
  {Mecking}},\ }\href {\doibase 10.1016/S0079-6425(02)00003-8} {\bibfield
  {journal} {\bibinfo  {journal} {Prog. Mater. Sci.}\ }\textbf {\bibinfo
  {volume} {48}},\ \bibinfo {pages} {171 } (\bibinfo {year}
  {2003})}\BibitemShut {NoStop}%
\bibitem [{\citenamefont {Klaum\"{u}nzer}\ \emph {et~al.}(2010)\citenamefont
  {Klaum\"{u}nzer}, \citenamefont {Maa\ss{}}, \citenamefont {Dalla~Torre},\
  and\ \citenamefont {L\"{o}ffler}}]{Klaumunzer2010}%
  \BibitemOpen
  \bibfield  {author} {\bibinfo {author} {\bibfnamefont {D.}~\bibnamefont
  {Klaum\"{u}nzer}}, \bibinfo {author} {\bibfnamefont {R.}~\bibnamefont
  {Maa\ss{}}}, \bibinfo {author} {\bibfnamefont {F.~H.}\ \bibnamefont
  {Dalla~Torre}}, \ and\ \bibinfo {author} {\bibfnamefont {J.~F.}\ \bibnamefont
  {L\"{o}ffler}},\ }\href {\doibase 10.1063/1.3309686} {\bibfield  {journal}
  {\bibinfo  {journal} {Appl. Phys. Lett.}\ }\textbf {\bibinfo {volume} {96}},\
  \bibinfo {eid} {061901} (\bibinfo {year} {2010})}\BibitemShut {NoStop}%
\bibitem [{\citenamefont {Rodney}\ and\ \citenamefont
  {Schuh}(2009)}]{Rodney2009}%
  \BibitemOpen
  \bibfield  {author} {\bibinfo {author} {\bibfnamefont {D.}~\bibnamefont
  {Rodney}}\ and\ \bibinfo {author} {\bibfnamefont {C.}~\bibnamefont {Schuh}},\
  }\href {\doibase 10.1103/PhysRevLett.102.235503} {\bibfield  {journal}
  {\bibinfo  {journal} {Phys. Rev. Lett.}\ }\textbf {\bibinfo {volume} {102}},\
  \bibinfo {pages} {235503} (\bibinfo {year} {2009})}\BibitemShut {NoStop}%
\bibitem [{\citenamefont {Mayr}(2006)}]{Mayr2006}%
  \BibitemOpen
  \bibfield  {author} {\bibinfo {author} {\bibfnamefont {S.~G.}\ \bibnamefont
  {Mayr}},\ }\href {\doibase 10.1103/PhysRevLett.97.195501} {\bibfield
  {journal} {\bibinfo  {journal} {Phys. Rev. Lett.}\ }\textbf {\bibinfo
  {volume} {97}},\ \bibinfo {pages} {195501} (\bibinfo {year}
  {2006})}\BibitemShut {NoStop}%
\bibitem [{\citenamefont {Sutton}\ and\ \citenamefont
  {Vitek}(1983)}]{Sutton1983}%
  \BibitemOpen
  \bibfield  {author} {\bibinfo {author} {\bibfnamefont {A.~P.}\ \bibnamefont
  {Sutton}}\ and\ \bibinfo {author} {\bibfnamefont {V.}~\bibnamefont {Vitek}},\
  }\href {\doibase 10.1098/rsta.1983.0020} {\bibfield  {journal} {\bibinfo
  {journal} {Philos Trans. Roy. Soc. London A}\ }\textbf {\bibinfo {volume}
  {309}},\ \bibinfo {pages} {1,37,55} (\bibinfo {year} {1983})}\BibitemShut
  {NoStop}%
\bibitem [{\citenamefont {Bernal}(1964)}]{Bernal1964}%
  \BibitemOpen
  \bibfield  {author} {\bibinfo {author} {\bibfnamefont {J.~D.}\ \bibnamefont
  {Bernal}},\ }\href {\doibase 10.1098/rspa.1964.0147} {\bibfield  {journal}
  {\bibinfo  {journal} {Proc. Roy. Soc. London A}\ }\textbf {\bibinfo {volume}
  {280}},\ \bibinfo {pages} {299} (\bibinfo {year} {1964})}\BibitemShut
  {NoStop}%
\bibitem [{\citenamefont {Spaepen}(1975)}]{Spaepen1975}%
  \BibitemOpen
  \bibfield  {author} {\bibinfo {author} {\bibfnamefont {F.}~\bibnamefont
  {Spaepen}},\ }\href {\doibase http://dx.doi.org/10.1016/0001-6160(75)90056-5}
  {\bibfield  {journal} {\bibinfo  {journal} {Acta Metall.}\ }\textbf {\bibinfo
  {volume} {23}},\ \bibinfo {pages} {729} (\bibinfo {year} {1975})}\BibitemShut
  {NoStop}%
\bibitem [{\citenamefont {Howe}(1997)}]{Howe1997}%
  \BibitemOpen
  \bibfield  {author} {\bibinfo {author} {\bibfnamefont {J.~M.}\ \bibnamefont
  {Howe}},\ }\href@noop {} {\emph {\bibinfo {title} {Interfaces in materials :
  atomic structure, thermodynamics and kinetics of solid vapor, solid liquid
  and solid solid interfaces}}}\ (\bibinfo  {publisher} {Wiley},\ \bibinfo
  {address} {New York},\ \bibinfo {year} {1997})\BibitemShut {NoStop}%
\bibitem [{\citenamefont {Johnson}\ and\ \citenamefont
  {Samwer}(2005)}]{Johnson2005}%
  \BibitemOpen
  \bibfield  {author} {\bibinfo {author} {\bibfnamefont {W.~L.}\ \bibnamefont
  {Johnson}}\ and\ \bibinfo {author} {\bibfnamefont {K.}~\bibnamefont
  {Samwer}},\ }\href {\doibase 10.1103/PhysRevLett.95.195501} {\bibfield
  {journal} {\bibinfo  {journal} {Phys. Rev. Lett.}\ }\textbf {\bibinfo
  {volume} {95}},\ \bibinfo {pages} {195501} (\bibinfo {year}
  {2005})}\BibitemShut {NoStop}%
\bibitem [{\citenamefont {Dasgupta}\ \emph {et~al.}(2013)\citenamefont
  {Dasgupta}, \citenamefont {Joy}, \citenamefont {Hentschel},\ and\
  \citenamefont {Procaccia}}]{Dasgupta2013}%
  \BibitemOpen
  \bibfield  {author} {\bibinfo {author} {\bibfnamefont {R.}~\bibnamefont
  {Dasgupta}}, \bibinfo {author} {\bibfnamefont {A.}~\bibnamefont {Joy}},
  \bibinfo {author} {\bibfnamefont {H.~G.~E.}\ \bibnamefont {Hentschel}}, \
  and\ \bibinfo {author} {\bibfnamefont {I.}~\bibnamefont {Procaccia}},\ }\href
  {\doibase 10.1103/PhysRevB.87.020101} {\bibfield  {journal} {\bibinfo
  {journal} {Phys. Rev. B}\ }\textbf {\bibinfo {volume} {87}},\ \bibinfo
  {pages} {020101} (\bibinfo {year} {2013})}\BibitemShut {NoStop}%
\bibitem [{\citenamefont {Chattoraj}\ \emph {et~al.}(2010)\citenamefont
  {Chattoraj}, \citenamefont {Caroli},\ and\ \citenamefont
  {Lema\^itre}}]{Chattoraj2010}%
  \BibitemOpen
  \bibfield  {author} {\bibinfo {author} {\bibfnamefont {J.}~\bibnamefont
  {Chattoraj}}, \bibinfo {author} {\bibfnamefont {C.}~\bibnamefont {Caroli}}, \
  and\ \bibinfo {author} {\bibfnamefont {A.}~\bibnamefont {Lema\^itre}},\
  }\href {\doibase 10.1103/PhysRevLett.105.266001} {\bibfield  {journal}
  {\bibinfo  {journal} {Phys. Rev. Lett.}\ }\textbf {\bibinfo {volume} {105}},\
  \bibinfo {pages} {266001} (\bibinfo {year} {2010})}\BibitemShut {NoStop}%
\bibitem [{\citenamefont {Megusar}\ \emph {et~al.}(1979)\citenamefont
  {Megusar}, \citenamefont {Argon},\ and\ \citenamefont {Grant}}]{Megusar1979}%
  \BibitemOpen
  \bibfield  {author} {\bibinfo {author} {\bibfnamefont {J.}~\bibnamefont
  {Megusar}}, \bibinfo {author} {\bibfnamefont {A.}~\bibnamefont {Argon}}, \
  and\ \bibinfo {author} {\bibfnamefont {N.}~\bibnamefont {Grant}},\ }\href
  {\doibase http://dx.doi.org/10.1016/0025-5416(79)90033-8} {\bibfield
  {journal} {\bibinfo  {journal} {Mater. Sci. Eng.}\ }\textbf {\bibinfo
  {volume} {38}},\ \bibinfo {pages} {63} (\bibinfo {year} {1979})}\BibitemShut
  {NoStop}%
\bibitem [{\citenamefont {Johnson}\ \emph {et~al.}(2007)\citenamefont
  {Johnson}, \citenamefont {Demetriou}, \citenamefont {Harmon}, \citenamefont
  {Lind},\ and\ \citenamefont {Samwer}}]{Johnson2007}%
  \BibitemOpen
  \bibfield  {author} {\bibinfo {author} {\bibfnamefont {W.~L.}\ \bibnamefont
  {Johnson}}, \bibinfo {author} {\bibfnamefont {M.~D.}\ \bibnamefont
  {Demetriou}}, \bibinfo {author} {\bibfnamefont {J.~S.}\ \bibnamefont
  {Harmon}}, \bibinfo {author} {\bibfnamefont {M.~L.}\ \bibnamefont {Lind}}, \
  and\ \bibinfo {author} {\bibfnamefont {K.}~\bibnamefont {Samwer}},\
  }\href@noop {} {\bibfield  {journal} {\bibinfo  {journal} {MRS Bulletin}\
  }\textbf {\bibinfo {volume} {32}},\ \bibinfo {pages} {644} (\bibinfo {year}
  {2007})}\BibitemShut {NoStop}%
\end{thebibliography}%


%

\end{document}